%% file: max.tex
\documentclass[11]{article}
\usepackage{amsmath,amsthm,amssymb, fullpage, graphicx}

\newtheorem{corollary}{Corollary}[section]

\newtheorem{definition}{Definition}[section]
\newtheorem{example}{Example}[section]
\newtheorem{lemma}{Lemma}[section]
\newtheorem{proposition}{Proposition}[section]
\newtheorem{theorem}{Theorem}[section]

\newcommand{\maxeq}{\textup{{\sc max-eq}}}
\newcommand{\mineq}{\textup{{\sc min-eq}}}
\newcommand{\algmax}{\textup{{\sc Alg-max-eq}}}

\title{\textbf{On Nash Dynamics of Matching Market Equilibria\\[.2in]}}

\author{Ning Chen$^*$ \and Xiaotie Deng$^\dag$\\[.1in] }

\date{\normalsize{}
$^*$Division of Mathematical Sciences,
Nanyang Technological University, Singapore. \\
Email: {\sf ningc@ntu.edu.sg}\\[.05in]
$^\dag$Department of Computer Science, University of Liverpool,
UK. \\
Email: {\sf xiaotie@liv.ac.uk} }

\begin{document}

\maketitle
\thispagestyle{empty}

\begin{abstract}
In this paper, we study the Nash dynamics of strategic interplays of
$n$ buyers in a matching market setup by a seller, the market maker.
Taking the standard market equilibrium approach, upon receiving
submitted bid vectors from the buyers, the market maker will decide
on a price vector to clear the market in such a way that each buyer
is allocated an item for which he desires the most (a.k.a., a market
equilibrium solution). While such equilibrium outcomes are not
unique, the market maker chooses one (\maxeq) that optimizes its own
objective --- revenue maximization. The buyers in turn change bids
to their best interests in order to obtain higher utilities in the
next round's market equilibrium solution.

This is an $(n+1)$-person game where buyers place strategic bids to
gain the most from the market maker's equilibrium mechanism. The
incentives of buyers in deciding their bids and the market maker's
choice of using the \maxeq\ mechanism create a wave of Nash dynamics
involved in the market. We characterize Nash equilibria in the
dynamics in terms of the relationship between \maxeq\ and \mineq\
(i.e., minimum revenue equilibrium), and develop convergence results
for Nash dynamics from the \maxeq\ policy to a \mineq\ solution,
resulting an outcome equivalent to the truthful VCG mechanism.

Our results imply revenue equivalence between \maxeq\ and \mineq,
and address the question that why short-term revenue maximization is
a poor long run strategy, in a deterministic and dynamic setting.
\end{abstract}

\nonumber

\newpage

\section{Introduction}



The Nash equilibrium paradigm in Economics has been based on a
rationality assumption that each individual will maximize its own
utility function. Making it further to a dynamic process where
multiple agents play interactively in a repeated game, the Nash
dynamics refers to a process where participants take turns to choose
a strategy to maximize their own utility function. We pose the
question of what impact of such strategic behavior of the seller, as
the market maker, can have on Nash dynamics of the buyers in a
matching market setting~\cite{SS}, where the seller has $m$ products
and there are $n$ unit-demand potential buyers with different
private values $v_{ij}$ for different items.

We base our consideration on the market equilibrium framework: The
market maker chooses a non-negative price vector and an allocation
vector; the outcome is called a {\em market competitive equilibrium}
if all items with positive prices are sold out and everyone gets his
maximum utility at the corresponding allocation, measured by the
difference between the buyer's value of the item and the charged
price, i.e., $v_{ij}-p_j$. To achieve its own objective, i.e.,
revenue maximization, the market maker is naturally to set all items
the highest possible prices so that there is still a supported
allocation vector that clears the market and maximizes everyone's
utility. This yields a seller-optimal outcome and is called a {\em
maximum competitive equilibrium}. Alternatively, at another extreme,
the outcome of the market can be buyers-optimal, i.e., a {\em
minimum competitive equilibrium} where all items have the lowest
possible prices to ensure market clearance and utility maximization.
Maximum and minimum equilibria represent the contradictory interests
of the two parties in a two-sided matching market at the two
extremes; and, as shown by Shapley and Shubik~\cite{SS}, both
equilibria always exist and their prices are unique.

Using market equilibrium as a mechanism for computing prices and
allocations yields desirable properties of efficiency and fairness.
The winners in the mechanism, however, will bid against the protocol
to reduce their payments; further, the losers may increase their
bids to be more competitive (to their best interests). The problem
becomes an $(n+1)$-person game played in turn between the seller and
the buyers where buyers make strategic moves to (mis)report their
bids for the items, in response to other buyers' bids and the
solution selected by the market maker.

It is well-known that the minimum equilibrium mechanism admits a
truthful dominant strategy for every buyer~\cite{Leo,DG}, resulting
in a solution equivalent to the VCG
protocol~\cite{vickrey,clarke,groves}. However, it is hardly
convincing a rational economic agent who dominates the market to
make a move that maximizes the collective returns of its buyers by
sacrificing its own revenue. Clearly, the Nash equilibrium paradigm
has always assumed that an economic agent is a utility maximizer;
for the market marker, it implies revenue maximization. Further,
success in bringing great revenue (in a short term) will bring a
large number of eyeballs and lead to positive externalities, which
could breed more success (in a long term). Indeed, quite a few
successful designs, like the generalized second price auction for
sponsored search markets~\cite{AGM} and the FCC spectrum auction for
licenses of electromagnetic spectra~\cite{Cra,Mil}, do not admit a
truthful dominant strategy.

Our interest is to find out the properties of the game between the
seller and the buyers (among which they compete with each other for
the items), especially with these two extremes of the solution
concepts --- their relationship proves to be significant in the
incentive analysis of the buyers in the market. In particular, we
are interested in the dynamics of the outcomes when the market maker
adopts the maximum equilibrium mechanism and the buyers adopt a
particular type of strategy, that of the best response strategy.
With the best response strategy, a buyer would announce his bids for
all items that would result in the maximum utility for himself,
under the fixed bids of other buyers and the maximum equilibrium
mechanism prefixed by the market maker. The best response strategy
varies depending on the circumstances of the market and has its
nature place in the analysis of dynamics of strategic interplays of
agents.

For the special case of single item markets where the market maker
has only one item to sell, maximum equilibrium corresponds to the
first price auction protocol and minimum equilibrium corresponds to
the second price auction protocol. The first price auction sells the
item to the buyer with the highest bid at a price equal to his bid
(i.e., the highest possible price so that everyone is happy with his
corresponding allocation).
Under a dynamic setting, buyers take turns in making their moves in
response to what they are encountered in the previous round(s). In a
simple setting when all buyers bid their true values initially, the
winner with the highest bid will immediately shade his bid down to
the smallest value so that he is still a winner. Such best response
bidding of the winner results in an outcome equivalent to the
Vickrey second price auction, which charges the winner at the
smallest possible price (i.e., the second highest bid) so that
everyone's utility is maximized.
The observation of convergence from the first price to the second
price goes hand-in-hand with the seminal revenue equivalence theory
in Bayesian analysis~\cite{vickrey,myerson}. Therefore, both the
Bayesian model and the dynamic best response model point to the same
unification result of the first price and the second price solution
concepts in the single item markets.

The story is hardly ending for the process of Nash dynamics with
multiple items being sold in the market, which illustrates a rich
structure that a single item market does not possess. First of all,
in a single item market, the best response decision of a buyer is
binary, that is, he decides either to or not to compete with the
highest bid of others for the item at a smallest cost. With multiple
items, however, a buyer needs to make his decision in the best
response over all items rather than a single item. In particular, he
may desire any item in a subset --- each of which brings him the
same maximum utility --- and lose interest for other items. The
buyer's best response, therefore, is in accordance with his
strategic bids for the items in {\em both} subsets. For example, it
is possible that one increases hid bid for one item but decreases
for another in a best response. This opens many possibilities that
would complicate the analysis; in particular, best responses of a
buyer are not unique. Indeed, as Example~\ref{example-not-converge}
shows, not all best responses guarantee convergence, again because
of the multiplicity in possibilities of bidding strategies.

Second, in a single item market, the submitted bids of buyers
illustrate a monotone property to the price of the item. That is,
when a buyer increases his bid, the price is always monotonically
non-decreasing. With multiple items, while it is true that the
prices of the items are still monotone with respect to the bids of
the losers, counter-intuitively, the prices are no longer monotone with
respect to the bids of the winners, as the following
Example~\ref{example-non-monotone} shows. This non-monotonicity
further requires extra efforts to analyze the multiplicity of best
response strategies and their properties.

\begin{example}[Non-monotonicity]\label{example-non-monotone}
There are three buyers $i_1,i_2,i_3$ and two items $j_1,j_2$ with
submitted bids $b_{i_1j_1}=2$, $b_{i_2j_1}=12$, $b_{i_2j_2}=14$ and
$b_{i_3j_2}=5$ (the bids of all unspecified pairs are zero). For the
given bid vector, $(p_{j_1},p_{j_2})=(3,5)$ is the maximum
equilibrium price vector, and assigning $j_1$ to $i_2$ and $j_2$ to
$i_3$ are supported allocations. If a winner $i_2$ increases his bid
of $j_2$ to $b'_{i_2j_2}=15$, then $(p'_{j_1},p'_{j_2})=(2,5)$ is
the maximum equilibrium price vector (with the same allocations)
where the price of $j_1$ is decreased from 3 to 2. Hence, the
maximum equilibrium prices are not monotone with respect to the
submitted bids (of the winners).
\end{example}

Finally, the maximum market equilibrium solution has no closed form
as in the single item case, and the convergence of the best
responses depends on a careful choice of the bidding strategy.
Recall that a major difficulty in analyzing the best responses is
that, in addition to bidding strategically for the items that bring
the maximum utility, buyers may have (almost arbitrarily) different
bidding strategies for those items that they are not interested in;
these decisions indeed play a critical role to the convergence of
the best responses. As not all best responses necessarily lead to
convergence (Example~\ref{example-not-converge}), we restrict to a
specific bidding strategy, called {\em aligned best response}, where
a bid vector of a buyer is called aligned if any allocation of an
item with positive bid brings him the maximum utility. The aligned
bidding strategy illustrates the preference of a buyer over all
items and is shown to be a best response
(Lemma~\ref{lemma-loser-best-response}
and~\ref{lemma-winner-best-response}). While the aligned best
response still does not have a monotone property in general (see
Figure~\ref{fig-syn-not-monotone}), the prices do exhibit a pattern
of monotonicity when the bids of the buyers have already been
aligned. Based on these properties, we show that the aligned best
response always converges and maximizes social welfare, summarized
by the following claim.

\medskip
\noindent \textbf{Theorem.} \textit{In the maximum competitive
equilibrium mechanism game, for any initial bid vector and any
ordering of the buyers, the aligned best response always converges.
Further, the allocation at convergence maximizes social welfare.}


\medskip
In addition to proving convergence, another important question is
that which Nash equilibrium to which the best response will
converge. In contrast with single item markets where Nash
equilibrium is essentially unique, in multi-item markets there can
be several Nash equilibria with
completely irrelevant price vectors (see
Appendix~\ref{appendix-example-nash}); this is another remarkable
difference between single item and multi-item markets. Despite of
the multiplicity of Nash equilibria, if we start with an aligned bid
vector (e.g., bid truthfully), the best response always converges to
one at a minimum competitive equilibrium, i.e., a VCG outcome.

\medskip
\noindent \textbf{Theorem.} \textit{Starting from an aligned bid
vector, the aligned best response of the maximum equilibrium
mechanism converges to a minimum competitive equilibrium at truthful
bidding.}

\subsection{Related Work and Motivation}

The study of competitive equilibrium in a matching market was
initiated by Shapley and Shubik in an assignment model~\cite{SS}.
They showed that maximum and minimum competitive equilibria always
exist and gave a simple linear program to compute one. Their results
were later improved to the models with general utility
functions~\cite{CK,Gale}. Leonard~\cite{Leo} and Demange and
Gale~\cite{DG} studied strategic behaviors in the market and proved
that the minimum equilibrium mechanism admits a truthful dominant
strategy for every buyer. Later, Demange, Gale and
Sotomayor~\cite{DGS} gave an ascending auction based algorithm that
converges to a minimum equilibrium. Our study focuses on Nash
dynamics in the matching market model; the convergence from maximum
equilibrium to minimum equilibrium implies revenue equivalence, and
addresses the question that why short-term revenue maximization is a
poor long run strategy in a dynamic framework.

The Nash dynamics of best responses has its nature place in the
analysis of interplays of strategic agents. In general,
characterizing equilibria of the dynamics is difficult or
intractable. There have been extensive studies in the literature for
some special settings, e.g., potential games~\cite{MS}, congestion
games~\cite{CS}, evolutionary games~\cite{HS}, concave
games~\cite{EMN}, correlated equilibrium~\cite{HM}, sink
equilibria~\cite{GMV}, and market equilibrium~\cite{WZ}, to name a
few.
Complexity issues have also been addressed in the analysis of best
responses~\cite{EB,FP,MS}. Recently, Nisan et al.~\cite{NSVZ} independently
considered best response dynamics in matching market and show a similar convergence result for 
running first price auctions for all items individually. To the best of our knowledge, our work is
the first to study best response dynamics in the maximum competitive
equilibrium mechanism.

Despite the motivation is mainly from theoretical curiosity, our
setting does capture some realistic applications, such as eBay
electronic market and sponsored search market, which have attracted
a lot research efforts in recent years.
The mechanism used in the sponsored search market is that of the
generalized second price (GSP) auction. Because GSP is not truthful
in general, a number of studies have focused on strategic
considerations of advertisers. Edelman et al.~\cite{EOS} and
Varian~\cite{varian} independently showed that certain Nash
equilibrium in the GSP auction derives the same revenue as the
well-known truthful VCG scheme. Cary et al.~\cite{CDE} showed that a
certain best response bidding strategy converges to the best Nash
equilibrium.
Recently, Leme and Tardos~\cite{PT} considered other possible Nash
equilibrium outputs and showed that the ratio between the worst and
best Nash equilibria is upper bounded by 1.618. These results
putting together illustrate a pretty complete overview of the
structure of strategic behaviors in GSP. Our results are not
directly about GSP but in a different way to reconfirm the revenue
equivalence:
While the search engine may adopt a different protocol with the goal
of revenue maximization (i.e., the maximum equilibrium mechanism),
with rational advertisers its overall revenue will eventually be the
same as in the VCG protocol.

Another widely studied problem related to our model is that of
spectrum markets. In designing the FCC spectrum auction protocol, a
multiple stage bidding process, proposed by Milgrom~\cite{Mil}, to
digest coordination, optimization and withdrawal, is adopted. It is
conducted in several stages to allow buyers to change their bids
when the seller announces the tentative prices of the licenses for
the winners. Therefore, it is created as an alternative game played
between the seller and the buyers as a whole. Our best response
analysis considers the dynamic aspect of the model and illustrates
the convergence of the dynamic process.

\medskip
{\em Organization.} We will first describe our model and
maximum/minimum competitive equilibrium (mechanism) in Section 2. In
Section 3, we define the aligned bidding strategy and show that it
is a best response and always converges. In Section 4, we
characterize Nash equilibria in the maximum equilibrium mechanism
game; and based on the characterization, we show that the maximum
equilibrium mechanism converges to a minimum equilibrium output. We
conclude our discussions in Section 5.

\section{Preliminaries}

We have a market with $n$ unit-demand buyers, where each buyer wants
at most one item, and $m$ indivisible items, where each item can be
sold to at most one buyer. We will denote buyers by $i$ and items by
$j$ throughout the paper. For every buyer $i$ and item $j$, there is
a {\em value} $v_{ij}\in [0, \infty)$, representing the maximum
amount that $i$ is willing to pay for item $j$. We will assume that
there are $m$ dummy buyers all with value zero for each item $j$,
i.e., $v_{ij} = 0$. This assumption is without loss of generality,
and implies that the number of items is always less than or equal to
the number of buyers, i.e., $m\le n$.

The outcome of the market is a tuple $(\mathbf{p},\mathbf{x})$,
where
\begin{itemize}
\item $\mathbf{p}=(p_1,\ldots,p_m)\ge 0$ is a \textit{price} vector, where $p_j$ is the price charged for item $j$;
\item $\mathbf{x}=(x_1,\ldots,x_n)$ is an \textit{allocation} vector, where $x_i$ is the item that $i$ wins. If $i$ does not win any items, denote $x_i=\emptyset$. Note that different buyers must win different items, i.e., $x_i\neq x_{i'}$ for any $i\neq i'$ if $x_i,x_{i'}\neq \emptyset$.
\end{itemize}
Given an output $(\mathbf{p},\mathbf{x})$, let
$u_i(\mathbf{p},\mathbf{x})$ denote the {\em utility} that $i$
obtains. We will assume that all buyers have quasi-linear utilities.
That is, if $i$ wins item $j$ (i.e., $x_i=j$), his utility is
$u_i(\mathbf{p},\mathbf{x})=v_{ij}-p_j$; if $i$ does not win any
item (i.e., $x_i=\emptyset$), his utility is
$u_i(\mathbf{p},\mathbf{x})=0$.

Buyers' preferences over items are according to their utilities ---
higher utility items are more preferable. We say that buyer $i$ {\em
(strictly) prefers} $j$ to $j'$ if $v_{ij}-p_j>v_{ij'}-p_{j'}$, is
{\em indifferent} between $j$ and $j'$ if
$v_{ij}-p_j=v_{ij'}-p_{j'}$, and {\em weakly prefers} $j$ to $j'$ if
$v_{ij}-p_j\ge v_{ij'}-p_{j'}$. In particular, a utility of zero,
$v_{ij}-p_j = 0$, means that $i$ is indifferent between buying item
$j$ at price $p_j$ and not buying anything at all; a negative
utility $v_{ij}-p_j <0$ means that the buyer strictly prefers to not
buy the item at price $p_j$.

We consider the following solution concept in this paper.

\begin{definition}{\sc (Competitive equilibrium)}
We say a tuple $(\mathbf{p},\mathbf{x})$ is a \textup{competitive
equilibrium} if (i) for any item $j$, $p_j=0$ if no one wins $j$ in
allocation $\mathbf{x}$, and (ii) for any buyer $i$, his utility is
maximized by his allocation at the given price vector. That is,
\begin{itemize}
\item if $i$ wins item $j$ (i.e., $x_i=j$), then $v_{ij}-p_j \ge 0$; and for every other item $j'$, $v_{ij}-p_j\ge v_{ij'}-p_{j'}$;
\item if $i$ does not win any item (i.e., $x_i=\emptyset$), then for every item $j$, $v_{ij}-p_j \leq
0$. (For notational simplicity, we write
$v_{i\emptyset}-p_{\emptyset}=0$.)
\end{itemize}
\end{definition}

The first condition above is an {\em efficiency} condition (i.e.,
market clearance), which says that all unallocated items are priced
at zero (or at some given reserve price). The assumption that there
is a dummy buyer for each item allows us to assume, without loss of
generality, that all items are allocated in an equilibrium. The
second is a {\em fairness} condition (i.e., envy-freeness), implying
that each buyer is allocated with an item that maximizes his utility
at these prices. That is, if $i$ wins item $j$, then $i$ cannot
obtain higher utility from any other item; and if $i$ does not win
any item, then $i$ cannot obtain a positive utility from any item.
Namely, all buyers are {\em happy} with their corresponding
allocations at the given price vector.

For any given matching market, Shapley and Shubik~\cite{SS} proved
that there always is a competitive equilibrium. Actually, what they
showed was much stronger --- there is the unique minimum
(respectively, maximum) equilibrium price vector, defined formally
as follows.

\begin{definition}{\sc (Minimum equilibrium \mineq\ and maximum equilibrium \maxeq)}
A price vector $\mathbf{p}$ is called a {\em minimum equilibrium
price vector} if for any other equilibrium price vector
$\mathbf{q}$, $p_j\le q_j$ for every item $j$. An equilibrium
$(\mathbf{p},\mathbf{x})$ is called a {\em minimum equilibrium}
(denoted by \mineq) if $\mathbf{p}$ is the minimum equilibrium price
vector. The maximum equilibrium price vector and a maximum
equilibrium (denoted by \maxeq) are defined similarly.
\end{definition}

For example, there are there buyers $i_1,i_2,i_3$ and one item; the
values of buyers are $v_{i_1} = 10, v_{i_2} = 5$ and $v_{i_3} = 2$.
Then $p=5$ and $p=10$ are the minimum and maximum equilibrium
prices, respectively. Allocating the item to the first buyer $i_1$,
together with any price $5\le p\le 10$, yields a competitive
equilibrium. When there is a single item, it can be seen that the
outcome of the minimum equilibrium and the maximum equilibrium
corresponds precisely to the ``second price auction" and the ``first
price auction", respectively.

Consider another example: There are there buyers $i_1,i_2,i_3$ and
two items $j_1,j_2$; the values of buyers are $v_{i_1j_1} = 10,
v_{i_1j_2} = 6$, $v_{i_2j_1} = 8, v_{i_2j_2} = 4$, and $v_{i_3j_1} =
3, v_{i_3j_2} = 2$. Then $(6,2)$ is the minimum equilibrium price
vector and $(8,4)$ is the maximum equilibrium price vector. Note
that the allocation vectors supported by the minimum or maximum
equilibrium price vector may not be unique. In this example, we can
allocate $j_1$ and $j_2$ arbitrarily to $i_1$ and $i_2$ to form an
equilibrium. Indeed, as Gul and Stacchetti~\cite{GulS} showed, if
both $\mathbf{p}$ and $\mathbf{q}$ are equilibrium price vectors and
$(\mathbf{p},\mathbf{x})$ is a competitive equilibrium, then
$(\mathbf{q},\mathbf{x})$ is an equilibrium as well.

\subsection{Maximum Equilibrium Mechanism Game}

In this paper, we will consider a maximum equilibrium as a
mechanism, that is, every buyer $i$ reports a {\em bid} $b_{ij}$ for
each item $j$ (note that $b_{ij}$ can be different from his true
value $v_{ij}$), and given reported bids from all buyers
$\mathbf{b}=(b_{ij})$, the {\em maximum equilibrium mechanism}
(again denoted by \maxeq) outputs a maximum equilibrium with respect
to $\mathbf{b}$. Let $\maxeq(\mathbf{b})$ denote the (maximum)
equilibrium returned by the mechanism.

Let $u_i(\mathbf{b})$ denote the utility that $i$ obtains in
$\maxeq(\mathbf{b})$. That is, if
$(\mathbf{p},\mathbf{x})=\maxeq(\mathbf{b})$, then
$u_i(\mathbf{b})=v_{ix_i}-p_{x_i}$ if $x_i\neq\emptyset$ and
$u_i(\mathbf{b})=0$ if $x_i=\emptyset$. Note that the (true) utility
of every buyer is defined according to his true valuations $v_{ij}$,
rather than the bids $b_{ij}$; and the ``equilibrium" output
$(\mathbf{p},\mathbf{x})$ is computed in terms of the bid vector
$\mathbf{b}$, rather than the true valuations (i.e., it is only
guaranteed that $b_{ix_i}-p_{x_i}\ge b_{ij}-p_{j}$ for any item
$j$). Therefore, the true utility of a buyer $i$ might not be
maximized at the corresponding allocation $x_i$ and the given price
vector $\mathbf{p}$. Further, the output
$(\mathbf{p},\mathbf{x})=\maxeq(\mathbf{b})$ might not even be a
real equilibrium with respect to the true valuations of the buyers.

Considering competitive equilibria as mechanisms defines a
multi-parameter setting and it is natural to consider strategic
behaviors of the buyers. While it is a dominant strategy for every
buyer to report his true valuations in the minimum equilibrium
mechanism (i.e., the mechanism outputs a \mineq\ for the given bid
vector)~\cite{DG}, the \maxeq\ mechanism does not in general admit a
dominant strategy. This is true even for the degenerated single item
case --- it is well-known that the first price auction is not
incentive compatible. Therefore, in the \maxeq\ mechanism, buyers
will behave strategically to maximize their utilities. In
particular, fixing bids of other buyers, buyer $i$ will naturally
place his bid vector $(b_{ij})_j$ to maximize his own utility; such
a vector is called a {\em best response}. The focus of the present
paper is to consider convergence of best response dynamics, i.e.,
buyers iteratively change their bids according to best responses
while all the others remain their previous bids unchanged. We say a
best response dynamics {\em converges} if it eventually reaches a
state where no buyer is willing to change his bid anymore; hence, it
arrives at a Nash equilibrium.

As our focus is on the convergence of best response dynamics, we
discretize the bidding space of every buyer to be a multiple of a
given arbitrarily small constant $\epsilon>0$, and assume that all
$v_{ij}$'s and $b_{ij}$'s are multiples of $\epsilon$. In practice,
$\epsilon$ can be, e.g., one cent or one dollar (or any other unit
number). This assumption is natural in the context of two-sided
markets with money transfers, and implies that if a buyer increases
his bid for an item, it must be by at least $\epsilon$. In addition,
we assume that the bid that every buyer submits for every item is
less than or equal to his true valuation, i.e., $b_{ij}\le v_{ij}$.
This assumption is rather mild because (i) bidding higher than the
true valuations carries the risk of a negative utility, and (ii) as
Lemma~\ref{lemma-loser-best-response} and
~\ref{lemma-winner-best-response} below show, for any given fixed
bids of other buyers, there always is a best response strategy for
every buyer to bid less than or equal to his true valuations.

\subsection{Computation of Maximum Equilibrium}

Shapley and Shubik~\cite{SS} gave a linear program to compute a
\mineq; their approach can be easily transformed to compute a
\maxeq\ by changing the objective to maximize the total payment.
Next we give a combinatorial algorithm to iteratively increase
prices to converge to a \maxeq. The idea of the algorithm is
important to our analysis in the subsequent sections.

\begin{definition}
[Demand graph] Given any given bid vector $\mathbf{b}$ and price
vector $\mathbf{p}$, its {\em demand graph}
$G(\mathbf{b},\mathbf{p})=(A,B;E)$ is defined as follows: $A$ is the
set of buyers and $B$ is the set of items, and $(i,j)\in E$ if
$b_{ij}\ge p_j$ and $b_{ij}- p_j\ge b_{ij'}-p_{j'}$ for any $j'$.
\end{definition}

In a demand graph $G(\mathbf{b},\mathbf{p})$, every edge $(i,j)$
represents that item $j$ gives maximal utility to buyer $i$,
presumed that his true valuations are given by $\mathbf{b}$. Note
that demand graph is uniquely determined by the given bid vector and
price vector, and is independent to any allocation vector. However,
an equilibrium allocation must be selected from the set of edges in
the demand graph. The following definition of alternating paths in
the demand graph is crucial to our analysis. (Recall that all items
are allocated out in an equilibrium.)

\begin{definition}{\sc ($\max$-alternating path)} Given any equilibrium
$(\mathbf{p},\mathbf{x})$ of a given bid $\mathbf{b}$, let
$G=G(\mathbf{b},\mathbf{p})$ be its demand graph. For any item $j$,
a path $(j=j_1,i_1,j_2,i_2,\ldots,j_\ell,i_\ell)$ in graph $G$ is
called a {\em $\max$-alternating path} if edges are in and not in
the allocation $\mathbf{x}$ alternatively, i.e., $x_{i_k}=j_k$ for
all $k=1,\ldots,\ell$. Denote by
$G_j^{\max}(\mathbf{b},\mathbf{p},\mathbf{x})$ (or simply
$G_j^{\max}$ when the parameters are clear from the context) the
subgraph of $G(\mathbf{b},\mathbf{p})$ (containing both buyers and
items including $j$ itself) reachable from $j$ through
$\max$-alternating paths with respect to $\mathbf{x}$. A
$\max$-alternating path $(j=j_1,i_1,j_2,i_2,\ldots,j_\ell,i_\ell)$
in $G_j^{\max}(\mathbf{b},\mathbf{p},\mathbf{x})$ is called {\em
critical} if $b_{i_\ell j_\ell}=p_{j_\ell}$.
\end{definition}

For any given bid vector $\mathbf{b}$, let $(\mathbf{p},\mathbf{x})$
be an arbitrary equilibrium. Note that for any buyer $i$ with
$x_i\neq \emptyset$, $(i,x_i)$ is an edge in the demand graph
$G(\mathbf{b},\mathbf{p})$. We consider the following recursive rule
to increase prices: For any item $j$, increase prices of all items
in $G_j^{\max}$ continuously with the same amount until one of the
following events occurs:
\begin{enumerate}
\item There is buyer $i\in G_j^{\max}$ such that $b_{ix_i}=p_{x_i}$ (note that $x_i\in G_j^{\max}$);
\item There is buyer $i\in G_j^{\max}$ and item $j'\notin G_j^{\max}$ such that $i$ obtains maximal utility from $j'$ as well; then we add
edge $(i,j')$ to the demand graph and update $G_j^{\max}$ for each
$j$.
\end{enumerate}
The process continues iteratively until we cannot increase the price
for any item. Denote the algorithm by \algmax. We have the following
result.

\begin{theorem}\label{theorem-alg-max-eq}
Starting from an arbitrary equilibrium $(\mathbf{p},\mathbf{x})$ for
the given bid vector $\mathbf{b}$, let $\mathbf{q}$ be the final
price vector in the above price-increment process \algmax. Together
with the original allocation vector $\mathbf{x}$,
$(\mathbf{q},\mathbf{x})$ is a \maxeq\ with respect to $\mathbf{b}$.
\end{theorem}

The above algorithm \algmax\ and theorem illustrate the idea of
defining $G_j^{\max}$ (and the corresponding $\max$-alternating
paths), summarized in the following corollary. (Note that if
$(j=j_1,i_1,j_2,i_2,\ldots,j_\ell,i_\ell)$ is a critical
$\max$-alternating path in
$G_{j}^{\max}(\mathbf{b},\mathbf{p},\mathbf{x})$, then the last pair
$(i_\ell,j_\ell)$ where $b_{i_\ell j_\ell}=p_{j_\ell}$ is the exact
reason that why we are not able to increase the price $p_j$ further
in the \algmax\ to derive a higher equilibrium price vector.)

\begin{corollary}\label{coro-critical-alter-path}
Given any bid vector $\mathbf{b}$ and
$(\mathbf{p},\mathbf{x})=\maxeq(\mathbf{b})$, for any item $j$,
there is a critical $\max$-alternating path in the subgraph
$G_{j}^{\max}(\mathbf{b},\mathbf{p},\mathbf{x})$.
\end{corollary}

\section{Best Response Dynamics}

For any given bid vector $\mathbf{b}$, assume that the \maxeq\
outputs $(\mathbf{p},\mathbf{x})$, i.e.,
$(\mathbf{p},\mathbf{x})=\maxeq(\mathbf{b})$. If
$(\mathbf{p},\mathbf{x})$ is not a Nash equilibrium, there is a
buyer who is able to obtain more utility by unilaterally changing
his bid. Such a buyer will therefore naturally choose a vector
(called a {\em best response}) to bid so that his utility is
maximized in the \maxeq\ mechanism, given fixed bids of all other
buyers.

Consider the following example: There are three buyers $i_1,i_2,i_3$
and two items $j_1,j_2$, with $b_{i_1j_1} = v_{i_1j_1} = 20$,
$b_{i_1j_2} = v_{i_1j_2} = 18$, and $b_{i_2j_1} = b_{i_2j_2} =
b_{i_3j_1} = b_{i_3j_2} = 10$. Then $\mathbf{p}=(12,10)$ is the
maximum equilibrium price and $i_1$ obtains utility 8. Consider a
scenario where $i_1$ changes his bid to
$b'_{i_1j_1}=b'_{i_1j_2}=15$; then the maximum equilibrium price
vector becomes $\mathbf{p'}=(10,10)$. Given the equilibrium price
vector $\mathbf{p'}$, however, there are two different equilibrium
allocations which give him different true utilities (where $i_1$
either wins $j_1$ with utility $v_{i_1j_1}-p'_{j_1}=10$ or wins
$j_2$ with utility $v_{i_1j_2}-p'_{j_2}=8$). Hence, different
selections of equilibrium allocations may lead to different true
utilities, which in turn will certainly affect best response
strategies of the buyers.

To analyze the best responses of the buyers, it is therefore
necessary to specify a framework about their belief on the resulting
equilibrium allocations. We will consider worst case analysis in
this paper, i.e., all buyers are risk-averse and always assume the
worst possible allocations when making their best
responses\footnote{When buyers report the same bids on some items, a
certain tie-breaking rule should be specified to decide an
equilibrium allocation. In the literature, ties are broken either by
an oracle access to the true valuations or by a given fixed order of
buyers~\cite{IKN,KKT}. In our worst-case analysis framework, we
actually do not need to specify a tie-breaking rule explicitly. It
essentially implies that (i) a buyer who changes his bid has the
lowest priority in tie-breaking, and (ii) any buyer who bids zero
cannot win the corresponding item at price zero due to the existence
of dummy buyers (i.e., there is no free lunch).}. The above bid
vector $b'_{i_1j_1}=b'_{i_1j_2}=15$, therefore, is not a best
response for $i_1$ since his utility in the worst allocation is only
8. In the worse-case analysis framework, the best responses of $i_1$
in the above example are given by $b'_{i_1j_1}=10+\epsilon$ and
$b'_{i_1j_2}\le 10$ where $i_1$ wins $j_1$ at price $10+\epsilon$
with utility $v_{i_1j_1}-(10+\epsilon) = 10-\epsilon$ (it can be
seen that there is no bid vector such that $i_1$ can obtain a
utility higher than or equal to 10 given fixed bids of the other two
buyers)\footnote{For any fixed bids of other buyers, the utility
that a risk-averse buyer obtains in a best response is always within
a distance of $\epsilon$ to that in a best possible allocation that
a risk-seeking buyer may obtain. Thus, the worst case analysis
(i.e., with risk-averse buyers) does provide a ``safe" equilibrium
allocation in which the corresponding utility is almost the same as
the maximal.}.

The above example further shows that best response strategies may
not be unique: While $i_1$ always wants to bid $10+\epsilon$ for
$j_1$, he can bid any value between 0 and 10 for $j_2$ to constitute
a best response. While the bids for those items in which a buyer is
not interested (i.e., $j_2$ in the above example) will not affect
the utility that the buyer obtains after his best response bidding,
they do affect the overall convergence of the best response
dynamics, as the following example shows.

\begin{figure*}[ht]
\begin{center}
\includegraphics[scale = 0.9]{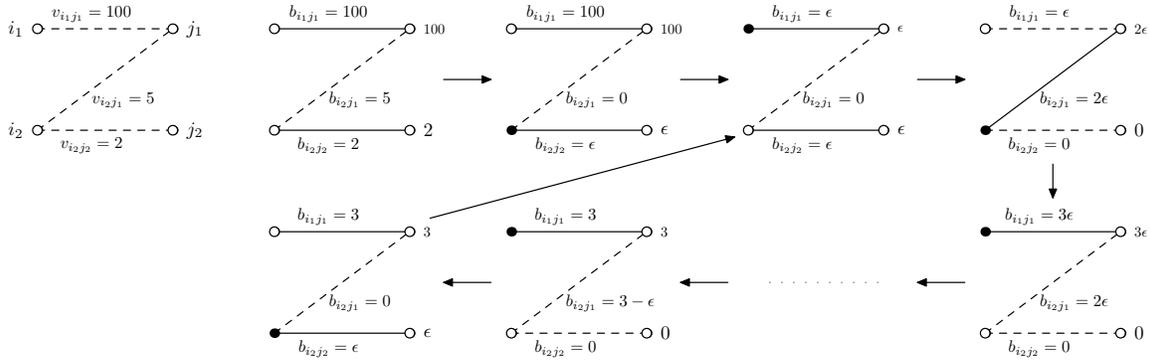}
\caption{A best response that does not
converge.}\label{fig-not-converge}
\end{center}
\end{figure*}

\begin{example}\label{example-not-converge}
{\sc (Non-convergence of a best response)} There are two buyers
$i_1,i_2$ and two items $j_1,j_2$, where $v_{i_1j_1}=100$,
$v_{i_1j_2}=0$, $v_{i_2j_1}=5$ and $v_{i_2j_2}=2$. (There still are
dummy buyers; we do not describe them explicitly.) Consider the
following specific best response strategy: Every buyer in the best
response always bids zero to those items that he is not interested
in. Since $v_{i_1j_2}=0$, we assume that $i_1$ always bids zero to
$j_2$. The process in Figure~\ref{fig-not-converge} shows that the
best response dynamics does not converge. (For all examples in the
paper, the black vertices on the left denote the buyers who change
their bids in the best response, the solid lines denote allocations,
and the numbers along with item vertices denote the maximum
equilibrium prices.)
\end{example}

\subsection{Best Response Strategy}

Given the non-convergence of a specific best response shown in
Example~\ref{example-not-converge}, one would ask if there are any
best response strategies that {\em always} converge for {\em any}
given instance. We consider the following bidding strategy.

\begin{lemma}\label{lemma-loser-best-response}
{\sc (Best response of losers)} For any given bid vector
$\mathbf{b}$, let $(\mathbf{p},\mathbf{x})=\maxeq(\mathbf{b})$.
Given fixed bids of other buyers, consider any buyer $i_0$ with
$x_{i_0}=\emptyset$; let $d_{i_0}=\max_{j}v_{i_0j}-p_j$.
\begin{itemize}
\item If $d_{i_0}\le 0$, a best response of $i_0$ is to bid the same vector $(b_{i_0j})_j$.

\item Otherwise, a best response of $i_0$ is given by vector $(b'_{i_0j})_j$, where $b'_{i_0j}=\max\{0,v_{i_0j}-d_{i_0}+\epsilon\}$. Further, the utility that $i_0$ obtains after the best response bidding is either $d_{i_0}$ or $d_{i_0}-\epsilon$.\footnote{Note that if $d_{i_0} > \epsilon$, then certainly $i_0$ improves his utility through the best response bidding. If $d_{i_0} = \epsilon$, however, it is possible that $i_0$ obtains the same utility $d_{i_0} - \epsilon = 0 = u_{i_0}(\mathbf{p},\mathbf{x})$ after the best response bidding. For such a case, $(b'_{i_0j})_j$ is still a best response and we assume that $i_0$ will continue to change his bid according to it. This assumption is without loss of generality as our focus is on the convergence of the best response strategy. Further, in practice, buyers usually do not have complete information about the market (e.g., bid vectors of others) and are unaware of the exact utility they will obtain if change their bids. Hence, continuing to bid the best response provides a safe strategy for the buyers to maximize their utilities.}
\end{itemize}
\end{lemma}

\begin{lemma}\label{lemma-winner-best-response}
{\sc (Best response of winners)} For any given bid vector
$\mathbf{b}$, let $(\mathbf{p},\mathbf{x})=\maxeq(\mathbf{b})$.
Given fixed bids of other buyers, consider any buyer $i_0$ with
$x_{i_0}\neq \emptyset$. Denote by $\mathbf{b}_{\not\ni i_0}$ the
bid vector derived from $ \mathbf{b}$ where $i_0$ changes his bid to
0 for all items. Let $(\mathbf{q},\mathbf{y}) =
\maxeq(\mathbf{b}_{\not\ni i_0})$ and $d_{i_0}=\max_jv_{i_0j}-q_j$.
\begin{itemize}
\item If $d_{i_0}\le 0$ or $u_{i_0}(\mathbf{p},\mathbf{x})=d_{i_0}$, a best response of $i_0$ is to bid the same vector $(b_{i_0j})_j$.

\item Otherwise, a best response of $i_0$ is given by vector $(b'_{i_0j})_j$, where $b'_{i_0j}=\max\{0,v_{i_0j}-d_{i_0}+\epsilon\}$. Further, the utility that $i_0$ obtains after the best response bidding is either $d_{i_0}$ or $d_{i_0}-\epsilon$.\footnote{Similar to the discussions for Lemma~\ref{lemma-loser-best-response}, if $u_{i_0}(\mathbf{p},\mathbf{x})< d_{i_0} - \epsilon$, then $i_0$ strictly improves his utility after the best response bidding; if $u_{i_0}(\mathbf{p},\mathbf{x}) = d_{i_0} - \epsilon$, it is possible that $i_0$ obtains the same utility $d_{i_0} - \epsilon$ after the best response bidding. Again for such a case, $(b'_{i_0j})_j$ is still a best response and we assume that $i_0$ will continue change his bid according to it.}
\end{itemize}
\end{lemma}

The best response $(b'_{i_0j})_j$ defined in the above two lemmas
has two remarkable properties: First, if $v_{i_0j}\ge v_{i_0j'}$
then $b'_{i_0j}\ge b'_{i_0j'}$; second, if $b'_{ij},b'_{ij'}>0$ then
$v_{i_0j}-b'_{i_0j}=v_{i_0j'}-b'_{i_0j'}$. Hence, the best response
$(b'_{i_0j})_j$ captures the preference of $i_0$ over all items. In
addition, given fixed bids of other buyers, the difference
$v_{i_0j}-b'_{i_0j}$ gives the maximal possible utility (up to a gap
of $\epsilon$) that $i_0$ is able to obtain from item $j$.

We say a bid vector $(b_{ij})_j$ {\em aligned} if for any
$b_{ij},b_{ij'}>0$, $v_{ij}-b_{ij}=v_{ij'}-b_{ij'}$. That is,
$(b_{ij})_j$ is derived from $(v_{ij})_j$ by moving the curve down
in parallel capped at 0. It can be seen that the bid vectors defined
in the above Lemma~\ref{lemma-loser-best-response} and
Lemma~\ref{lemma-winner-best-response} are aligned. Hence, we will
refer such bidding strategy by {\em aligned best response}. In the
following, unless stated otherwise, all best responses are aligned
according to these two lemmas. The following
Figure~\ref{fig-converge} shows the convergence of
Example~\ref{example-not-converge} according to the aligned best
response when both buyers bid their true values initially (as in
Figure~\ref{fig-not-converge}).
\begin{figure}[ht]
\begin{center}
\includegraphics[scale = 0.9]{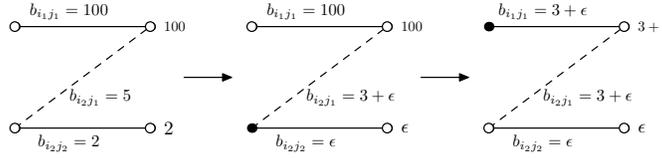}
\caption{Convergence of Example~\ref{example-not-converge} by the
aligned best response.}\label{fig-converge}
\end{center}
\end{figure}

\begin{figure*}[ht]
\begin{center}
\includegraphics[scale = 0.85]{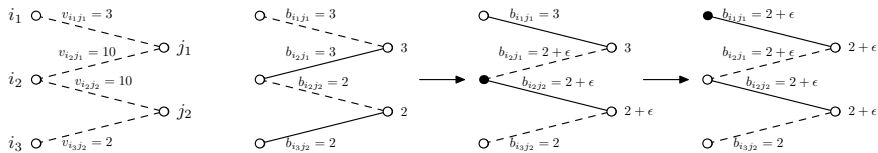}
\caption{Non-monotonicity of the aligned best
response.}\label{fig-syn-not-monotone}
\end{center}
\end{figure*}

\subsection{Properties of the Best Response}

Example~\ref{example-non-monotone} shows that maximum equilibrium
prices may not be monotone with respect to the submitted bid
vectors. Such non-monotonicity still occurs even for the aligned
best response of the winners defined in
Lemma~\ref{lemma-winner-best-response}. In the example of the
Figure~\ref{fig-syn-not-monotone}, when a winner $i_2$ changes his
bid, the price of $j_2$ is increased from 2 to $2+\epsilon$ in the
maximum equilibrium. Hence, the maximum equilibrium prices may not
be monotonically decreasing with respect to the aligned best
responses of the winners.

However, as the following claim shows, we do have monotonicity for
the maximum equilibrium prices given a certain condition. This
property is crucial in the analysis of convergence.

\begin{lemma}\label{lemma-monotone}
When a loser makes a best response bidding (by
Lemma~\ref{lemma-loser-best-response}), the maximum equilibrium
prices will not decrease. On the other hand, when a winner, who has
already made at least one aligned best response bidding, makes a
best response bidding (by Lemma~\ref{lemma-winner-best-response}),
the maximum equilibrium prices will not increase.
\end{lemma}

We have the following corollary.

\begin{corollary}\label{coro-best-response}
Given a bid vector $\mathbf{b}$ and
$(\mathbf{p},\mathbf{x})=\maxeq(\mathbf{b})$, any loser $i_0$ is
willing to make a best response bidding only if there is an item $j$
such that $v_{i_0j}>p_j$. For any winner $i_0$ who has already made
a best response bidding, the following claims hold:
\begin{itemize}
\item $i_0$ is willing to make a best response bidding $\mathbf{b'}$ only if the price of item $x_{i_0}$ will decrease in $\maxeq(\mathbf{b}_{\not\ni i_0})$ (defined in Lemma~\ref{lemma-winner-best-response}).
\item Let $(\mathbf{q},\mathbf{y})=\maxeq(\mathbf{b}_{\not\ni i_0})$ and $(\mathbf{p'},\mathbf{x'})=\maxeq(\mathbf{b'})$, then $(\mathbf{p'},\mathbf{x})$ is a \maxeq\ for $\mathbf{b'}$ as well. Hence, we can assume without loss of generality that when $i_0$ places a best response bidding, the allocation remains the same.
\end{itemize}
\end{corollary}

\subsection{Convergence of the Best Response Dynamics}

In this subsection we will show that the aligned best response
defined in Lemma~\ref{lemma-loser-best-response}
and~\ref{lemma-winner-best-response} always converges. In the rest
of this subsection we will assume that all buyers have already made
a best response bidding; this assumption is without loss of
generality since our goal is to prove convergence. Hence, the
results established in the last subsection can be applied directly.

\begin{proposition}\label{prop-winner-winner}
Given a bid vector $\mathbf{b}$ and
$(\mathbf{p},\mathbf{x})=\maxeq(\mathbf{b})$, assume that a winner
$i_0$ first makes a best response bidding.
\begin{itemize}
\item If next another winner $i'_0$ makes a best response bidding, then $i_0$ will not change his bid in the best
response.
\item If next a loser $i'_0$ makes a best response bidding, then as long as $i_0$ is still a winner after $i'_0$'s bid, he will not change his bid in the best response.
\end{itemize}
\end{proposition}

Similar to Proposition~\ref{prop-winner-winner}, we have the
following claim for the best response of losers.

\begin{proposition}\label{prop-loser-loser}
Given a bid vector $\mathbf{b}$ and
$(\mathbf{p},\mathbf{x})=\maxeq(\mathbf{b})$, assume that a loser
$i_0$ first makes a best response bidding and becomes a winner.
\begin{itemize}
\item If next another winner $i'_0$ makes a best response bidding, then $i_0$ will not change his bid in the best
response.
\item If next another loser $i'_0$ makes a best response bidding, then as long as $i_0$ is still a winner after $i'_0$'s bid, he will not change his bid in the best response.
\end{itemize}
\end{proposition}

The above two propositions imply the following corollary.

\begin{corollary}\label{coro-winner-loser}
Given a bid vector $\mathbf{b}$ and
$(\mathbf{p},\mathbf{x})=\maxeq(\mathbf{b})$,
\begin{itemize}
\item if a winner $i_0$ makes a best response bidding, then unless $i_0$ becomes a
loser (due to the best responses of others), he will not change his
bid in the best response;

\item if a loser $i_0$ makes a best response bidding and
becomes a winner, then unless $i_0$ becomes a loser again (due to
the best responses of others), he will not change his bid in the
best response.
\end{itemize}
\end{corollary}

We are now ready to prove our main theorem.

\begin{theorem}\label{theorem-convergence}
In the maximum competitive equilibrium mechanism game, for any given
initial bid vector and any ordering of the buyers, the aligned best
response defined in Lemma~\ref{lemma-loser-best-response}
and~\ref{lemma-winner-best-response} always converges.
\end{theorem}
\begin{proof}
By Corollary~\ref{coro-winner-loser}, after all buyers have already
made a best response bidding, only losers are willing to make best
responses to become a winner and winners are willing to make best
responses only if they become a loser. By
Lemma~\ref{lemma-monotone}, the prices will be monotonically
non-decreasing when losers make best responses. Hence, the process
eventually terminates.
\end{proof}

The above theorem only says that the aligned best response is
guaranteed to converge in finite steps; indeed it may take a
polynomial of $\max_{ij}v_{ij}/\epsilon$ steps to converge. This is
true even for the simplest setting with one item and two buyers of
the same value $v_{1}=v_{2}$: The two buyers with initial bids zero
keep increasing their bids by $2\epsilon$ one by one to beat each
other until the price reaches to $v_1=v_2$. However, in most
applications like advertising markets, the value of
$\max_{ij}v_{ij}$ is rather small. Further, to guarantee efficient
convergence, we may set $\epsilon$ to be sufficiently large so that
the rally between the winners and losers is fast (e.g., in most
ascending auctions, the minimum increment of a bid is scaled
according to the expected value of $\max_{ij}v_{ij}$).

\section{Characterization of Nash Dynamics: From {\sc max-eq} to {\sc min-eq}}

Theorem~\ref{theorem-convergence} in the previous section shows that
the aligned best response always converges for any given initial bid
vector. However, it does not answer the questions of at which
condition(s) the dynamics of the best response stops (i.e., reaches
to a stable state), and how the stable state looks like. We will
answer these questions in this section.

\begin{theorem}\label{lemma-stable-max=min}
For any Nash equilibrium bid vector $\mathbf{b}$ of the \maxeq\
mechanism, let $(\mathbf{p},\mathbf{x})=\mineq(\mathbf{b})$ and
$(\mathbf{q},\mathbf{y})=\maxeq(\mathbf{b})$. Then for any item $j$,
we have either $p_j=q_j$ or $p_j+\epsilon=q_j$. That is, the \maxeq\
and \mineq\ price vectors differ by at most $\epsilon$, i.e.,
$\maxeq(\mathbf{b})\approx \mineq(\mathbf{b})$.
\end{theorem}

The above characterizations apply to {\em all} Nash equilibria in
the \maxeq\ mechanism, and gives a necessary condition that when a
bid vector forms a Nash equilibrium. Note that the other direction
of the claim does not hold. For example, there are one item and two
buyers with the same true value $v_1=v_2=10$. In the bid vector
$\mathbf{b}$ with $b_1=b_2=5$, we have
$\maxeq(\mathbf{b})=\mineq(\mathbf{b})$. However, $\mathbf{b}$ is
not a Nash equilibrium as the loser in the $\maxeq(\mathbf{b})$ can
increase his bid to win the item.

In general, there can be multiple Nash equilibria for a given
market. To compare different Nash equilibria, we use a universal
benchmark --- the solution given by the truthful \mineq\ mechanism,
i.e., $(\mathbf{p}^*,\mathbf{x}^*)=\mineq(\mathbf{v})$, where
$\mathbf{v}$ is the true valuation vector. This is equivalent to the
outcome of the second price auction in single item markets and the
VCG mechanism for the general multi-item markets. In
Appendix~\ref{appendix-example-nash}, we give two examples to show
that, for a given Nash equilibrium $\mathbf{b}$ with
$(\mathbf{p},\mathbf{x})=\maxeq(\mathbf{b})$, there could be no
fixed relation between the price vectors $\mathbf{p}$ and
$\mathbf{p}^*$, i.e., it can be either $\mathbf{p}\gg\mathbf{p}^*$
or $\mathbf{p}\ll\mathbf{p}^*$. This is a remarkable difference
between single item and multi-item markets --- in the former we
always have $\mathbf{p}\approx\mathbf{p}^*$ (i.e.,
$||\mathbf{p}-\mathbf{p}^*||\le \epsilon$).

However, if we initially start with an aligned bid vector for all
buyers (e.g., bid truthfully), then there is a strong connection
between the \maxeq\ price vector at a converged Nash equilibrium and
$(\mathbf{p}^*,\mathbf{x}^*)$: The two price vectors are ``almost"
identical, up to a gap of $\epsilon$. We summarize our results in
the following theorem.

\begin{theorem}\label{lemma-aligned-price=p*}
For any Nash equilibrium bid vector $\mathbf{b}$ converged from the
aligned best response, starting from an aligned bid vector for all
buyers, let $(\mathbf{p},\mathbf{x})=\maxeq(\mathbf{b})$ and
$(\mathbf{p}^*,\mathbf{x}^*)=\mineq(\mathbf{v})$, where
$\mathbf{v}=(v_{ij})$ is the true valuation vector. Then we have
\begin{itemize}
\item either $p_j=p^*_j$ or $p_j=p^*_j+\epsilon$ for all items;
\item $\maxeq(\mathbf{b})\approx \mineq(\mathbf{b})\approx \mineq(\mathbf{v})$.
\end{itemize}
\end{theorem}

We comment that if we start with an {\em arbitrary} bid vector, as
long as all buyers bid at least one aligned best response in the
dynamics, the convergence result from \maxeq\ to \mineq\ still
holds. In particular, this includes the case when all buyers bid
very low at the beginning. In addition, for any given bid vector
$\mathbf{b}$, if the \maxeq\ mechanism outputs $\maxeq(\mathbf{b}) =
\mineq(\mathbf{v})$, it is not necessary that $\mathbf{b}$ is a Nash
equilibrium. For example, it is possible that the bids of losers are
quite small and therefore winners still want to decrease their bids
to pay less.

Finally, we characterize allocations at all Nash equilibria. We say
an allocation $\mathbf{x}$ {\em efficient} if $\sum_{i}v_{ix_i}$ is
maximized, i.e., social welfare is maximized. It is well-known that
all allocations in competitive equilibria are efficient at truthful
bidding. This can be easily generalized to say that for any given
bid vector $\mathbf{b}$, any equilibrium allocation $\mathbf{x}$
maximizes $\sum_{i}b_{ix_i}$. We have the following result, which
says that any Nash equilibrium converged from the aligned best
response actually maximizes the real social welfare with respect to
$\mathbf{v}$.

\begin{theorem}\label{theorem-social-welfare}
For any Nash equilibrium bid vector $\mathbf{b}$ of the \maxeq\
mechanism converged from the aligned best response (starting from an
arbitrary bid vector), let
$(\mathbf{p},\mathbf{x})=\maxeq(\mathbf{b})$. Then $\mathbf{x}$ is
an efficient allocation that maximizes social welfare.

\end{theorem}

We note that the above claim does not hold for general Nash
equilibria. For example, buyers $i_1$ and $i_2$ are interested in
item $j_1$ at values $v_{i_1j_1}=v$ and $v_{i_2j_1}=v+\epsilon$.
Then allocating the item to $i_1$ at price $v$ is a Nash equilibrium
for the bid vector $b_{i_1j_1}=b_{i_2j_1}=v$ (since the second buyer
$i_2$ still obtains a non-positive utility even if he can win the
item at a higher bid). This allocation does not maximize social
welfare, and the example can be easily generalized to arbitrary
number of items with large deficiency. In the aligned best response
bidding, however, $i_2$ will continue to bid $v+\epsilon$ to win the
item by Lemma~\ref{lemma-loser-best-response}, which leads to an
efficient allocation.

\section{Concluding Remarks}

In this paper, we analyze the dynamics of best responses of the
maximum competitive equilibrium mechanism in a matching market.
While a best response strategy may not necessarily converge, we show
that a specific bidding strategy, aligned best response, always
converges to a Nash equilibrium. The outcome at such a Nash
equilibrium is actually a minimum equilibrium given truthful bidding
if we start with an aligned bid vector. In other words, our results
show that maximum equilibrium converges to minimum equilibrium,
which is a reminiscence of the convergence of dynamic bidding from
first price auction to second price auction in a single item market.

In our discussions, we assume that all buyers have complete
information for the market, including, say, submitted bid vectors of
others, market prices and their corresponding allocation. We note
that the first one, submitted bids of other buyers, in not necessary
in the dynamics: The best response of losers defined in
Lemma~\ref{lemma-loser-best-response} applies automatically, and the
best response of winners defined in
Lemma~\ref{lemma-winner-best-response} can be implemented by a
two-step strategy without requiring extra information in the market.




\input{appendix-max.tex}

\end{document}

%% file: appendix-max.tex
\appendix

\section{Proof of Theorem~2.1}\label{appendix-proof:theorem-alg-max-eq}

\begin{lemma}\label{lemma-basic}
Let $(\mathbf{p},\mathbf{x})$ and $(\mathbf{p'},\mathbf{x'})$ be any
two equilibria for the given bid vector $\mathbf{b}$. Let
$T=\{j~|~p_j < p'_j\}$ and $S=\{i~|~x'_i\in T\}$ be the subset of
buyers who win items in $T$ in the equilibrium
$(\mathbf{p'},\mathbf{x'})$. Then all buyers in $S$ win all items in
$T$ in the equilibrium $(\mathbf{p},\mathbf{x})$ as well. In
particular, this implies that if $j$ is allocated to a dummy buyer
in $(\mathbf{p},\mathbf{x})$ (i.e., essentially no one wins $j$),
then its price is zero in all equilibria.
\end{lemma}
\begin{proof}
Note that since $p'_j > p_j\ge 0$ for any $j\in T$, all items in $T$
must be sold out in equilibrium $(\mathbf{p'},\mathbf{x'})$; hence,
$|S|=|T|$ and all buyers in $S$ win all items in $T$ in
$(\mathbf{p'},\mathbf{x'})$. Consider any buyer $i\in S$, let
$j'=x'_i\in T$. For any item $j\notin T$, since $p_j\ge p'_j$, we
have
\[b_{ij'}-p_{j'} > b_{ij'}-p'_{j'} \ge b_{ij}-p'_{j} \ge b_{ij}-p_{j} \]
where the second inequality follows from the fact that
$(\mathbf{x}',\mathbf{p'})$ is an equilibrium. Hence, buyer $i$
always strictly prefers item $j'$ to all other items not in $T$ at
price vector $\mathbf{p}$, which implies that his allocation $x_i\in
T$.
\end{proof}

\medskip
\noindent {\em Proof of Theorem~\ref{theorem-alg-max-eq}.} By the
rule of increasing prices, since the initial
$(\mathbf{p},\mathbf{x})$ is an equilibrium, all buyers are always
satisfied with their respective allocations in the course of the
algorithm. Further, it can be seen that any item that is allocated
to a dummy buyer (i.e., it has initial price $p_j=0$) will never
have its price increased. Hence, the final output
$(\mathbf{q},\mathbf{x})$ is an equilibrium as well.

Assume that $(\mathbf{x}^*,\mathbf{q^*})$ is a \maxeq\ for the given
bid vector $\mathbf{b}$; we know that $q_{j}\le q^*_{j}$ for all
items. Let $T=\{j~|~q_j < q^*_j\}$ and $S=\{i~|~x^*_i\in T\}$ be the
subset of buyers who win items in $T$ in the \maxeq\ equilibrium. By
Lemma~\ref{lemma-basic}, we know that all buyers in $S$ win all
items in $T$ in equilibrium $(\mathbf{q},\mathbf{x})$ as well.
Assume that $T\neq \emptyset$; and consider any $j_1\in T$ and the
subgraph $G_{j_1}^{\max}(\mathbf{b},\mathbf{q},\mathbf{x})$
reachable from $j_1$ in the demand graph of the final output
$(\mathbf{q},\mathbf{x})$.

We claim that all items in $G_{j_1}^{\max}$ are in $T$. Otherwise,
consider an item $j_\ell\in G_{j_1}^{\max}\setminus T$ and the
$\max$-alternating path $(j_1,i_1,j_2,i_2,\ldots,i_{\ell-1},j_\ell)$
defining $j_\ell$ to be in $G_{j_1}^{\max}$. Assume without loss of
generality that $j_\ell$ is the first item on the path which is not
in $T$, i.e., $j_{\ell-1}\in T$ and $j_\ell\notin T$. since
$j_{\ell-1} = x_{i_\ell-1}\in T$, by Lemma~\ref{lemma-basic}, we
have $j^*\triangleq x^*_{i_{\ell-1}}\in T$. Hence,
\begin{eqnarray*}
b_{i_\ell-1j_\ell}-q^*_{j_\ell} &\ge& b_{i_\ell-1j_\ell}-q_{j_\ell} \\
&=& b_{i_\ell-1j_\ell-1}-q_{j_\ell-1}\\
&\ge& b_{i_\ell-1j^*}-q_{j^*} \\
&>& b_{i_\ell-1j^*}-q^*_{j^*}
\end{eqnarray*}
which contradicts to the fact that $i_{\ell-1}$ obtains his
utility-maximized item in the \maxeq.

Since all items in $G_{j_1}^{\max}$ have their prices increased,
again by Lemma~\ref{lemma-basic}, all items in $G_{j_1}^{\max}$ are
sold out. Therefore, at the end of the algorithm when reaching to
$(\mathbf{q},\mathbf{x})$, we should still be able to increase
prices for items in $G_{j_1}^{\max}$, which is a contradiction. That
is, $T=\emptyset$ and $(\mathbf{q},\mathbf{x})$ is a \maxeq. \hfill
$\square$

\section{Proofs in Section~3}\label{appendix-proof-converge}

\subsection{Proof of Lemma~\ref{lemma-loser-best-response}}

\begin{proof}
Given fixed bids of other buyers, consider any bid vector
$(b^*_{i_0j})_j$ of buyer $i_0$. Denote the resulting bid vector by
$\mathbf{b}^*$, and let
$(\mathbf{p^*},\mathbf{x^*})=\maxeq(\mathbf{b}^*)$. A basic
observation is that no matter what bid that $i_0$ submits, everyone
other than $i_0$ is still happy with the original equilibrium
$(\mathbf{p},\mathbf{x})$. Hence, if $b^*_{i_0j}\le p_j$ for any
$j$, then $(\mathbf{p},\mathbf{x})$ is still a \maxeq\ for
$\mathbf{b}^*$. Otherwise, let $j_1\in \arg\max_j
\{b^*_{i_0j}-p_j\}$. Consider subgraph
$G^{\max}_{j_1}(\mathbf{b},\mathbf{p},\mathbf{x})$ and its critical
$\max$-alternating path $(j_1,i_1,j_2,i_2,\ldots,j_\ell,i_{\ell})$
with respect to $\mathbf{x}$, where $x_{i_k}=j_k$ for
$k=1,\ldots,\ell-1$ and the pair $i_{\ell}$ and $j_\ell$ is the
reason that $p_{j_1}$ cannot be increased in
$(\mathbf{p},\mathbf{x})$ by the algorithm (i.e., $b_{i_\ell
j_\ell}=p_{j_\ell}$). Consider reallocating each $j_k$ to $i_{k-1}$
for $k=2,\ldots,\ell$. By the definition of $\max$-alternating path,
we know that all these buyers are still happy with their new
allocations. Further, we reallocate $j_1$ to $i_0$; then $i_0$
obtains his utility-maximized item at price $\mathbf{p}$ under bid
vector $\mathbf{b}^*$. This new allocation, together with the price
vector $\mathbf{p}$, constitutes an equilibrium. In both cases,
$\mathbf{p}$ is an equilibrium price vector under bid vector
$\mathbf{b}^*$. Hence, the price of every item in
$\maxeq(\mathbf{b}^*)$ is larger than or equal to $p_j$, i.e.,
$p^*_j\ge p_j$.

We next analyze the best response of $i_0$ defined in the statement
of the claim. If $v_{i_0j}\le p_j$ for any $j$, then $i_0$ cannot
get a positive utility from any item at price vector $\mathbf{p}$,
as well as $\mathbf{p}^*$. Hence, bidding the original (losing)
price vector is a best response strategy. It suffices to consider
there is an item $j$ such that $v_{i_0j} > p_j$ and the best
response strategy $(b'_{i_0j})_j$ described in the second part of
the statement (denoted by $\mathbf{b}'$).

Let $T=\{j~|~v_{i_0j}-p_j=d_{i_0}\}$; by the assumption, $T\neq
\emptyset$. It can be seen that for any $j\in T$,
$b'_{ij}-p_j=\epsilon$; and any $j\notin T$, $b'_{ij}-p_{j} \le
v_{i_0j}-d_{i_0}+\epsilon - p_j < d_{i_0}-d_{i_0} + \epsilon =
\epsilon$, i.e., $b'_{ij}-p_{j} \le 0$. That is, given bid vector
$\mathbf{b}'$ and price vector $\mathbf{p}$, $i_0$ always desires
those items in $T$. Consider any item $j_0\in T$, by the same
reassignment argument described above, we can reallocate $j_0$ to
$i_0$, as well as a few other reallocations through a critical
$\max$-alternating path, to derive an equilibrium
$(\mathbf{p},\mathbf{x}')$, where $\mathbf{x}'$ is the corresponding
new allocation. Note that $(\mathbf{p},\mathbf{x}')$ may not be a
\maxeq. Consider subgraph
$G^{\max}_{j_0}(\mathbf{b}',\mathbf{p},\mathbf{x'})$; we ask whether
$p_{j_0}$ can be increased further by the algorithm \algmax\ to get
a \maxeq.

If the answer is `no', then $(\mathbf{p},\mathbf{x}')$ is indeed a
\maxeq\ under bid vector $\mathbf{b}'$ (it can be shown that the
price of any other item cannot be increased as well), and the
utility that $i_0$ obtains satisfies
\begin{eqnarray*}
u_{i_0}(\mathbf{p},\mathbf{x}') &=& v_{i_0j_0}-p_{j_0}=d_{i_0} \\
&=& \max_j v_{i_0j}-p_j \ge \max_j v_{i_0j}-p^*_j \ge
u_{i_0}(\mathbf{p^*},\mathbf{x^*})
\end{eqnarray*}

If the answer is `yes', then we can increase prices of all items in
$G^{\max}_{j_0}(\mathbf{b}',\mathbf{p},\mathbf{x'})$ by $\epsilon$,
which gives an equilibrium $(\mathbf{p'},\mathbf{x'})$, where
$p'_j=p_j+\epsilon$ if $j\in
G^{\max}_{j_0}(\mathbf{b}',\mathbf{p},\mathbf{x'})$ and $p'_j=p_j$
otherwise. Further, $(\mathbf{p'},\mathbf{x'})$ is a \maxeq\ under
bid vector $\mathbf{b}'$ since the price of $j_0$ is tight with the
bid of $i_0$, i.e., $b'_{i_0j_0}=p'_{j_0}$ (again, prices of other
items cannot be increased). Hence, the utility that $i_0$ obtains is
\[u_{i_0}(\mathbf{p'},\mathbf{x'})=v_{i_0j_0}-p'_{j_0}=v_{i_0j_0}-p_{j_0}-\epsilon=d_{i_0}-\epsilon\]
Next we consider the utility that $i_0$ obtains in the equilibrium
$(\mathbf{p^*},\mathbf{x^*})$ with bid vector $\mathbf{b}^*$. Note
that all other buyers bid the same values in $\mathbf{b},
\mathbf{b}'$ and $\mathbf{b^*}$. Let $j_1 \triangleq x^*_{i_0}$;
then
\[u_{i_0}(\mathbf{p^*},\mathbf{x^*})= v_{i_0j_1}-p^*_{j_1}\le v_{i_0j_1}-p_{j_1} \le d_{i_0}\]
If one of the above inequalities is strict, then
$u_{i_0}(\mathbf{p^*},\mathbf{x^*})<d_{i_0}$. That is,
$u_{i_0}(\mathbf{p^*},\mathbf{x^*})\le d_{i_0}-\epsilon =
u_{i_0}(\mathbf{p'},\mathbf{x'})$, which implies that $\mathbf{b}'$
is a best response strategy. It remains to consider the case when
all inequalities are tight, i.e.,
$u_{i_0}(\mathbf{p^*},\mathbf{x^*})=d_{i_0}$ and
$p^*_{j_1}=p_{j_1}$; in this case, we have $j_1\in T$. Consider the
following two cases regarding the relation between $j_0$ and $j_1$.

\begin{itemize}
\item If $j_0 = j_1$, consider the subgraph
$G^{\max}_{j_1}(\mathbf{b}^*,\mathbf{p}^*,\mathbf{x}^*)$. Since the
price of $j_1$ cannot be increased, there is a critical
$\max$-alternating path inside
$P=(j_1,i_1=i_0,j_2,i_2,\ldots,j_\ell,i_{\ell})$ where
$x^*_{i_k}=j_{k}$ and $b^*_{i_{\ell}j_\ell}=p^*_{j_\ell}$. Since
\[b^*_{i_0j_2}-p^*_{j_2}=b^*_{i_0j_1}-p^*_{j_1}=b^*_{i_0j_1}-p_{j_1}\]
we know that $j_2\in T$ and $p^*_{j_2}=p_{j_2}$ (otherwise,
$v_{i_0j_2}-p^*_{j_2}<d_{i_0}$, then in the worst allocation the
utility of $i_0$ is less than $d_{i_0}$). We claim that all items in
$P$ have $p^*_j=p_j$.  Otherwise, consider the first item $j_k$
where $p^*_{j_k}>p_{j_k}$. Note that $k\ge 3$; then we have
\begin{eqnarray*}
b'_{i_{k-1}j_{k-1}}-p_{j_{k-1}} &=& b^*_{i_{k-1}j_{k-1}}-p^*_{j_{k-1}} \\
&=& b^*_{i_{k-1}j_{k}}-p^*_{j_{k}} \\
&<& b^*_{i_{k-1}j_{k}}-p_{j_{k}} \\
&=& b'_{i_{k-1}j_{k}}-p_{j_{k}}
\end{eqnarray*}
Hence, $x'_{i_{k-1}}\in T^*\triangleq\{j~|~p^*_j>p_j\}$ and
$x^*_{i_{k-1}}=j_{k-1}\notin T^*$. This implies that there must be a
buyer $i$ such that $x'_i\notin T^*$ and $x^*_i\in T^*$, which is
impossible since $i$ does not get his utility-maximized item in
$(\mathbf{p},\mathbf{x'})$. Therefore, essentially $P$ defines a
critical $\max$-alternating path in
$G^{\max}_{j_0}(\mathbf{b}',\mathbf{p},\mathbf{x}')$.

\item If $j_0\neq j_1$, starting from $P'=(j_0,i_0,j_1)$, we expand the
path $P'$ through the following rule: if the current last edge is
$(i_k,j_{k+1})$, expand $(j_{k+1},i_{k+1})$ if
$x'_{i_{k+1}}=j_{k+1}$; if the current last edge is $(j_k,i_k)$,
expand $(i_k,j_{k+1})$ if $x^*_{i_k}=j_{k+1}$. The process stops
when there is no more item or buyer to expand; denote the final path
by $P'=(j_0,i_0,j_1,i'_1,j'_2,i'_2,\ldots,j_\ell,i'_\ell)$. Note
that edges in $P'$ are in $\mathbf{x}'$ and $\mathbf{x}^*$
alternatively. Further, since $i'_1$ wins $j_1$ in $\mathbf{x}'$ and
$j'_2$ in $\mathbf{x}^*$, plus the fact that $p^*_{j_1}=p_{j_1}$, we
have $p^*_{j'_2}=p_{j'_2}$; then since $i'_2$ wins $j'_2$ in
$\mathbf{x}'$ and $j'_3$ in $\mathbf{x}^*$, we have
$p^*_{j'_3}=p_{j'_3}$; the argument inductively implies that for
every item $j\in P'$, $p^*_j=p_j$. At the end of path $P'$,
$i'_\ell$ does not win any item in $\mathbf{x}^*$, which implies
that $b'_{i'_\ell j'_\ell}=p_{j'_\ell}$. If we consider path $P'$ in
$G^{\max}_{j_0}(\mathbf{b}',\mathbf{p},\mathbf{x}')$, the above
arguments show that it is actually a critical $\max$-alternating
path.
\end{itemize}
Hence, in both cases we cannot increase price $p_{j_0}$ in the
equilibrium $(\mathbf{p},\mathbf{x}')$, which contradicts to our
assumption that the answer is `yes'.
\end{proof}

\subsection{Proof of Lemma~\ref{lemma-winner-best-response}}

\begin{proof}
Given fixed bids of other buyers, consider any bid vector
$(b^*_{i_0j})_j$ of buyer $i_0$. Denote the resulting bid vector by
$\mathbf{b}^*$, and let
$(\mathbf{p^*},\mathbf{x^*})=\maxeq(\mathbf{b}^*)$. Consider the two
equilibria $(\mathbf{q},\mathbf{y})$ and
$(\mathbf{p^*},\mathbf{x^*})$ in the following virtual scenario:
$i_0$ first bids 0 and loses in the \maxeq\
$(\mathbf{q},\mathbf{y})$ and then bids according to $\mathbf{b}^*$
yielding a new \maxeq\ $(\mathbf{p^*},\mathbf{x^*})$. By a similar
argument as the proof of the above lemma, we know that $q_j\le
p^*_j$ for all items. In particular, the argument applies to the
case when $\mathbf{b}^* = \mathbf{b}$, hence $q_j\le p_j$.

Since the bid of any buyer is always less than or equal to his true
value, we have
\[u_{i_0}(\mathbf{p},\mathbf{x}) = v_{i_0x_{i_0}} - p_{x_{i_0}} \ge b_{i_0x_{i_0}} - p_{x_{i_0}} \ge 0\]
If $u_{i_0}(\mathbf{p},\mathbf{x})=d_{i_0}$, then certainly $i_0$
cannot obtain more utility when bidding $\mathbf{b}^*$ since
\[u_{i_0}(\mathbf{p}^*,\mathbf{x}^*) = v_{i_0x^*_{i_0}} - p^*_{x^*_{i_0}} \le v_{i_0x^*_{i_0}} - q_{x^*_{i_0}} \le d_{i_0} = u_{i_0}(\mathbf{p},\mathbf{x})\]
If $d_{i_0}\le 0$, then
\[v_{i_0x_{i_0}} - p_{x_{i_0}} \le v_{i_0x_{i_0}} - q_{x_{i_0}}\le 0\]
and
\[u_{i_0}(\mathbf{p^*},\mathbf{x^*})=v_{i_0x^*_{i_0}} - p^*_{x^*_{i_0}} \le v_{i_0x^*_{i_0}} - q_{x^*_{i_0}}\le 0\]
Hence, $u_{i_0}(\mathbf{p},\mathbf{x})=0\ge
u_{i_0}(\mathbf{p^*},\mathbf{x^*})$, which implies that bidding the
original vector $(b_{i_0j})_j$ is a best response strategy.

It remains to consider $d_{i_0}>u_{i_0}(\mathbf{p},\mathbf{x}) \ge
0$ and analyze the best response $(b'_{i_0j})_j$, denote by
$\mathbf{b}'$, defined in the statement of the claim. Consider a
virtual scenario where $i_0$ first bids 0 and loses in the
equilibrium $(\mathbf{q},\mathbf{y})$. By the above
Lemma~\ref{lemma-loser-best-response} for loser's best response, we
know that $\mathbf{b}'$ is a best response strategy for $i_0$ in the
virtual scenario and $u_{i_0}(\mathbf{b'})\ge d_{i_0}-\epsilon \ge
u_{i_0}(\mathbf{p},\mathbf{x})$. Since bidding according to
$\mathbf{b}^*$ is a specific strategy for $i_0$, we have
$u_{i_0}(\mathbf{b'})\ge u_{i_0}(\mathbf{p^*},\mathbf{x^*})$. These
two inequalities together imply that bidding according to
$\mathbf{b}'$ is a best response strategy for $i_0$.
\end{proof}

\subsection{Proof of Lemma~\ref{lemma-monotone}}

\begin{proof}
The first part of the claim follows directly from the proof of
Lemma~\ref{lemma-loser-best-response}, thus we will only prove the
second part. Consider bid vector $\mathbf{b}$ and equilibrium
$(\mathbf{p},\mathbf{x})=\maxeq(\mathbf{b})$, bid vector
$\mathbf{b}_{\not\ni i_0}$ (derived from $\mathbf{b}$ where a winner
$i_0$ bids 0 for all items) and equilibrium
$(\mathbf{q},\mathbf{y})=\maxeq(\mathbf{b}_{\not\ni i_0})$, and best
response $\mathbf{b}'$, all defined in the statement of
Lemma~\ref{lemma-winner-best-response}. If $i_0$ does not change his
bid, then the maximum equilibrium prices remain the same. Hence, in
the following we assume that $i_0$ changes his bid according to
$\mathbf{b}'$ and analyze the relation between $\mathbf{p}$ and
$\mathbf{p}'$, where
$(\mathbf{p'},\mathbf{x'})=\maxeq(\mathbf{b}')$. What we need to
show is that $\mathbf{p}'\le \mathbf{p}$.

By the proof of Lemma~\ref{lemma-winner-best-response}, we know that
$q_j\le p_j$ for all items. If $\mathbf{q}=\mathbf{p}$, i.e.,
$q_j=p_j$ for all items, let $j^*=\arg\max_j v_{ij}-q_j$. Note that
$d_{i_0}=v_{i_0j^*}-q_{j^*} > 0$. Since $(\mathbf{p},\mathbf{x})$ is
an equilibrium with respect to $\mathbf{b}$, we have
$b_{i_0x_{i_0}}-p_{x_{i_0}} \ge b_{i_0j^*}-p_{j^*}$ and
$b_{i_0x_{i_0}}\neq 0$. Since $i_0$ has already made a best response
bidding (either as a loser or a winner), his bids for different
items are aligned, i.e., $v_{i_0x_{i_0}} - b_{i_0x_{i_0}} \ge
v_{i_0j^*} - b_{i_0j^*}$. Thus,
\[u_{i_0}(\mathbf{p},\mathbf{x}) = v_{i_0x_{i_0}}-p_{x_{i_0}} \ge v_{i_0j^*}-p_{j^*} = v_{i_0j^*}-q_{j^*} = d_{i_0}\]
Hence, $i_0$ already obtains his maximally possible utility and will
not change his bid. In the following, we assume $\mathbf{q}\le
\mathbf{p}$ and there is an item such that $q_j<p_j$.

Let $T=\{j~|~q_j<p_j\}$ and $S=\{i~|~x_i\in T\}$ be the set of
buyers who win items in $T$ in $(\mathbf{p},\mathbf{x})$. We claim
that $i_0\in S$. Otherwise, all buyers in $S$ win all items in $T$
in $(\mathbf{q},\mathbf{y})$ as well with positive utilities and
they strictly prefer their corresponding allocations to those items
that are not in $T$. Hence, we can increase the prices of all items
in $T$ by $\epsilon$ to derive another equilibrium; this contradicts
to the fact that $(\mathbf{q},\mathbf{y})$ is a \maxeq. Therefore,
there is exactly one buyer $i^*\notin S$ who wins an item in $T$ in
$(\mathbf{q},\mathbf{y})$ given bid vector $\mathbf{b}_{\not\ni
i_0}$, i.e., $y_{i^*}\in T$. Next when $i_0$ changes his bid vector
according to $\mathbf{b'}$, by the proof of
Lemma~\ref{lemma-loser-best-response}, the new allocation vector
$\mathbf{x'}$, together with the given price vector $\mathbf{q}$,
constitutes an equilibrium. Let $j_0=x'_{i_0}$; since
\begin{eqnarray*}
v_{i_0j_0}-q_{j_0} &=& \max_jv_{i_0j}-q_j = d_{i_0} \\
&>& u_{i_0}(\mathbf{p},\mathbf{x}) \\
&=& v_{i_0x_{i_0}}-p_{x_{i_0}} \\
&\ge& v_{i_0j_0}-p_{j_0}
\end{eqnarray*}
we must have $j_0\in T$. That is, when $i_0$ ``joins the market
again" from $\mathbf{b}_{\not\ni i_0}$ to $\mathbf{b}'$, he grabs an
item in $T$ ``again". Since every buyer in $S\setminus \{i_0\}$
obtains a positive utility for his corresponding allocation in $T$
in $(\mathbf{q},\mathbf{y})$ and strictly prefers it to those items
that are not in $T$, he has to win an item in $T$ in
$(\mathbf{q},\mathbf{x'})$. Hence, the only buyer who is kicked out
of winning an item in $T$ is $i^*$. That is, buyers in $S$ ``again"
win all items in $T$. Finally, without loss of generality, we can
assume that in $\mathbf{x'}$ all items that are not in $T$ have the
same allocations as $\mathbf{x}$; this still keeps an equilibrium
and will not affect the computation of the maximum equilibrium price
vector. Our argument above can be summarized as below:
\begin{center}
\begin{tabular}{|l|r|c|}\hline
bid vector & equilibrium \ \ \ & winners for items in $T$ \\\hline
\hspace{.25in} $\mathbf{b}$  &  \maxeq\ $(\mathbf{p},\mathbf{x})$ &
$S\setminus\{i_0\}\cup \{i_0\}$  \\ \hline \hspace{.25in}
$\mathbf{b}_{\not\ni i_0}$  &  \maxeq\ $(\mathbf{q},\mathbf{y})$   &
$S\setminus\{i_0\}\cup \{i^*\}$  \\ \hline
  &  $(\mathbf{q},\mathbf{x}')$   &    \\  \cline{2-2}
\hspace{.25in} \raisebox{1.5ex}[0pt]{$\mathbf{b}'$}  &  \maxeq\
$(\mathbf{p'},\mathbf{x'})$   &
\raisebox{1.5ex}[0pt]{$S\setminus\{i_0\}\cup \{i_0\}$}  \\ \hline
\end{tabular}
\end{center}

Finally we consider running the algorithm \algmax\ on equilibrium
$(\mathbf{q},\mathbf{x}')$ to derive the maximum equilibrium
$(\mathbf{p'},\mathbf{x'})$. By the proof of
Lemma~\ref{lemma-loser-best-response}, we have either $p'_j=q_j$ or
$p'_j=q_j+\epsilon$. To the end of proving $p'_j\le p_j$, it remains
to show that it is impossible that $p'_j=q_j+\epsilon=p_j+\epsilon$
for all items; assume otherwise that there is such an item $j_\ell$
(this implies that $j_\ell\notin T$). Then again by the proof of
Lemma~\ref{lemma-loser-best-response}, the price of $j_\ell$ is
increased (from $\mathbf{q}$ to $\mathbf{p}'$) through a
$\max$-alternating path
$P=(j_0,i_0,j_1,i_1,\ldots,j_{\ell-1},i_{\ell-1},j_\ell)$ in the
subgraph $G^{\max}_{j_0}(\mathbf{b}',\mathbf{q},\mathbf{x'})$. That
is, for $k=0,1,\ldots,\ell-1$, $x'_{i_k}=j_k$ and
$v_{i_kj_k}-q_{j_k}=v_{i_kj_{k+1}}-q_{j_{k+1}}$. Consider subgraph
$G^{\max}_{j_\ell}(\mathbf{b},\mathbf{p},\mathbf{x})$, since
$(\mathbf{p},\mathbf{x})=\maxeq(\mathbf{b})$, we cannot increase
$p_{j_\ell}$ and there is a critical $\max$-alternating path $P'$ in
$G^{\max}_{j_\ell}(\mathbf{b},\mathbf{p},\mathbf{x})$. Further, it
can be seen that all items in $P'$ are not in $T$ (since every buyer
in $P'$ strictly prefers the corresponding allocation in
$\mathbf{x}$ to those items in $T$ at price vector $\mathbf{p}$).
Since all items that are not in $T$ have the same allocations in
$\mathbf{x}$ and $\mathbf{x}'$, we can expand path $P$ through $P'$,
which gives a critical $\max$-alternating path in
$G^{\max}_{j_0}(\mathbf{b}',\mathbf{q},\mathbf{x'})$. Therefore, we
cannot increase the price of any item in $P\cup P'$, including
$j_\ell$. This contradicts to our assumption that
$p'_{j_\ell}=q_{j_\ell}+\epsilon$; hence the lemma follows.
%
\end{proof}

\subsection{Proof of Corollary~\ref{coro-best-response}}

\begin{proof}
The claims for best responses follow directly from
Lemma~\ref{lemma-loser-best-response}
and~\ref{lemma-winner-best-response}, and the proof of
Lemma~\ref{lemma-monotone}. For the last claim, note that if
$(\mathbf{q},\mathbf{x})$ is an equilibrium for $\mathbf{b'}$, then
by the algorithm \algmax\ to increase prices to derive the maximum
equilibrium price vector $\mathbf{p'}$, $(\mathbf{p'},\mathbf{x})$
is a \maxeq\ for $\mathbf{b'}$. Recall in the proof of
Lemma~\ref{lemma-monotone}, $(\mathbf{q},\mathbf{x}')$ is an
equilibrium of $\mathbf{b'}$ and all items that are not in $T$ have
the same allocations as $\mathbf{x}$, where $T=\{j~|~q_j<p_j\}$.
Thus, it remains to show that all items in $T$ have the same
allocations in $\mathbf{x}$ and $\mathbf{x'}$. Since all items in
$T$ are allocated to the same subset of buyers $S$ in both
$\mathbf{x}$ and $\mathbf{x'}$, for any buyer $i\in S$, we have
\begin{eqnarray*}
u_i(\mathbf{p},\mathbf{x}) &=& b_{ix_i}-p_{x_i} \ge b_{ix'_i}-p_{x'_i} \\
u_i(\mathbf{q},\mathbf{x'}) &=& b_{ix'_i}-q_{x'_i} \ge
b_{ix_i}-q_{x_i}, \ \textup{if} \ i\neq i_0
\end{eqnarray*}
and
\begin{eqnarray*}
u_{i_0}(\mathbf{q},\mathbf{x'}) &=& b'_{i_0x'_{i_0}}-q_{x'_{i_0}} \ge b'_{i_0x_{i_0}}-q_{x_{i_0}} \\
&\Longrightarrow&  b_{i_0x'_{i_0}}-q_{x'_{i_0}} \ge
b_{i_0x_{i_0}}-q_{x_{i_0}}
\end{eqnarray*}
(The last inequality is because of the following argument: Assume
that $b'_{i_0x'_{i_0}} = b_{i_0x'_{i_0}} - \delta$. Since $i_0$ has
already made a best response bidding, his bid vector over different
items has already been aligned. By the best responses defined in
Lemma~\ref{lemma-loser-best-response}
and~\ref{lemma-winner-best-response}, we have $v_{i_0x'_{i_0}} -
b_{i_0x'_{i_0}} \le v_{i_0x'_{i_0}} - b'_{i_0x'_{i_0}}$. Hence
$\delta\ge 0$ and $b'_{i_0x_{i_0}} =\max\{0,b_{i_0x_{i_0}} -
\delta\} \ge b_{i_0x_{i_0}} - \delta$. Therefore,
$b_{i_0x'_{i_0}}-q_{x'_{i_0}} = \delta + b'_{i_0x'_{i_0}}
-q_{x'_{i_0}} \ge \delta + b'_{i_0x_{i_0}}-q_{x_{i_0}} \ge
b_{i_0x_{i_0}}-q_{x_{i_0}}$.) If items in $T$ are allocated in
different ways in $\mathbf{x}$ and $\mathbf{x'}$, then there are
$T'\subseteq T$ and $S'\subseteq S$ such that items in $T'$ are
allocated to buyers in $S'$ in $\mathbf{x}$ and $\mathbf{x'}$ which
forms an augmenting path. Adding the above two inequalities
(regarding $u_i(\mathbf{p},\mathbf{x})$ and
$u_i(\mathbf{q},\mathbf{x'})$) for all buyers in $S'$ gives
$\sum_{j\in T'}p_j+q_j \ge \sum_{j\in T'}p_j+q_j$. Hence, all
inequalities are tight, which implies that every buyer gets the same
utility from $x_i$ and $x'_i$. Therefore, we can change the
allocation of items in $T'$ according to $\mathbf{x}$, which still
gives an equilibrium.
\end{proof}

\subsection{Proof of Proposition~\ref{prop-winner-winner}}

\begin{proof}
Assume that $(\mathbf{q},\mathbf{y})=\maxeq(\mathbf{b}_{\not\ni
i_0})$ and $(\mathbf{p'},\mathbf{x})=\maxeq(\mathbf{b'})$, where
$\mathbf{b'}$ is the resulting bid vector after $i_0$ makes his best
response bidding (by Corollary~\ref{coro-best-response}, we can
assume that the equilibrium allocation is the same for $\mathbf{b}$
and $\mathbf{b'}$). Further, by the proof of
Corollary~\ref{coro-best-response}, we know that
$(\mathbf{q},\mathbf{x})$ is an equilibrium for $\mathbf{b'}$. In
the two allocations $\mathbf{x}$ and $\mathbf{y}$, consider the
following alternating path:
\[P=(i_0,j_0=x_{i_0},i_1,j_1,\ldots,i_\ell,j_\ell,i_{\ell+1})\]
where $i_k$ wins $j_k$ in $\mathbf{x}$ and $i_{k+1}$ wins $j_k$ in
$\mathbf{y}$, and $x_{i_{\ell+1}}=\emptyset$ (this implies, in
particular, $b_{i_{\ell+1}j_\ell}=q_{j_\ell}$). Further, we have
\[b_{i_{k+1}j_k}-q_{j_k} = b_{i_{k+1}j_{k+1}}-q_{j_{k+1}}, \ \textup{for} \ k=0,\ldots,\ell\]
We will analyze the relation between $\mathbf{q}$ and the price
vector after $i'_0$ makes his best response bidding to show the
desired result.

\begin{itemize}
\item First consider the next best response is made by another winner $i'_0$. After $i'_0$ makes his best response bidding denote the resulting bid vector by $\mathbf{b}''$), by Corollary~\ref{coro-best-response}, we know that all winners have the same allocations $\mathbf{x}$. Further, the above set of equations still holds for all buyers in $P$ for $\mathbf{b}''$. This is because, if $i'_0$ is one of them, say $i_{k+1}=i'_0$, then both $b_{i_{k+1}j_k}$ and $b_{i_{k+1}j_{k+1}}$ are reduced by the same amount to derive $b''_{i_{k+1}j_k}$ and $b''_{i_{k+1}j_{k+1}}$; so we still have $b''_{i_{k+1}j_k}-q_{j_k} = b''_{i_{k+1}j_{k+1}}-q_{j_{k+1}}$. Hence, the price of any item $j$ in path $P$, as well as those that give the maximal utility to $i_0$, cannot be smaller than $q_j$ in $\maxeq(\mathbf{b}'')$ (otherwise, $i_{\ell+1}$ has to be a winner).

    Therefore, when $i_0$ bids zero for all items after $i'_0$ makes his best response bidding, the price of item $j_0$ cannot be smaller than $q_{j_0}$ in a $\maxeq(\mathbf{b''}_{\not\ni i_0})$ as reallocating items according to $P$ where $i_{\ell+1}$ becomes a winner gives an equilibrium allocation. By Corollary~\ref{coro-best-response}, which says that every winner obtains the same item after his best response bidding, we know that the best response of $i_0$ is to bid the same vector, i.e., do not change his bid.

\item Next consider the next best response is made by a loser $i'_0$; let $\mathbf{b}''$ be the resulting bid vector. By the proof of Lemma~\ref{lemma-loser-best-response}, let $(\mathbf{p'},\mathbf{x'})$ be an equilibrium of $\mathbf{b''}$, where $\mathbf{p'}$ is the maximum equilibrium price of $\mathbf{b'}$ defined above and $\mathbf{x'}$ is derived from $\mathbf{x}$ (the allocation before $i'_0$'s bid) through an alternating path: \[P'=(i'_0, j'_1, i'_1,\ldots,j'_{r},i'_{r})\] where $i'_k$ wins $j'_k$ in $\mathbf{x}$ and $i'_k$ wins $j'_{k+1}$ in $\mathbf{x'}$, and $i'_r$ does not win in $\mathbf{x'}$. (Note that to derive a \maxeq\ for $\mathbf{b''}$, we still need to verify if the price of $j'_1$ can be increased in the subgraph $G_{j'_1}^{\max}(\mathbf{b''},\mathbf{p'},\mathbf{x'})$.)

    Assume that $i_0$ is still a winner after $i'_0$'s bid; note that the item that $i_0$ wins in $\mathbf{x'}$ can be either the one defined above according to $P'$ or the same item $j_0=x_{i_0}$. Next we consider the setting when $i_0$ bids zero for all items, i.e., $\mathbf{b''}_{\not\ni i_0}$, for the two possibilities respectively.

    If it is the former, i.e., $i'_k=i_0, j'_k=j_0$ and $x'_{i_0}=j'_{k+1}$, then the price of $j'_k$ and $j'_{k+1}$ cannot be smaller than $p'_{j'_{k}}$ (which is at least $q_{j'_{k}}$) and $p'_{j'_{k+1}}$ (which is at least $q_{j'_{k+1}}$) respectively in a $\maxeq(\mathbf{b''}_{\not\ni i_0})$ as we are able to reallocate items $j'_{k+1},\ldots,j'_r$ back to $i'_{k+1},\ldots,i'_r$, respectively, where $j'_r$ becomes a winner. Since $d_{i_0}$ defined in Lemma~\ref{lemma-winner-best-response} will not increase, the best response of $i_0$ is to bid the same vector, i.e., do not change his bid.

    If it is the latter, i.e., $i_0$ wins the same item $j_0$ in $\mathbf{x}$ and $\mathbf{x'}$, we claim that the price of $j_0$ cannot be smaller than $q_{j_0}$ in a $\maxeq(\mathbf{b''}_{\not\ni i_0})$. This is because: (i) We can reallocate items according to $P$ such that $i_{\ell+1}$ becomes a winner. (ii) If any item in $P'$ appears in $P$ or $i'_0=i_{\ell+1}$ (i.e., $i'_0$ is the last buyer on path $P$), similar to the above arguments, we can reallocate items according $P$ and $P'$ such that $i'_r$ becomes a winner. (Note that for the last case $i'_0=i_{\ell+1}$, since $b_{i'_0j_\ell}=b_{i_{\ell+1}j_\ell}=q_{j_\ell}$ and the bids of $i'_0$ have already been aligned, $j_\ell$ will give the maximal utility for $i_0$ when its price is $q_{j_\ell}$.)
\end{itemize}
\end{proof}

\subsection{Proof of Proposition~\ref{prop-loser-loser}}

\begin{proof}
An equivalent way to consider the best response of $i_0$ is that he
first bids 0 for all items (which yields the same \maxeq\
$(\mathbf{p},\mathbf{x})$), then bids according to the best response
of winners defined in Lemma~\ref{lemma-winner-best-response}. Then
we can apply the same argument as
Proposition~\ref{prop-winner-winner} to get the desired result.
\end{proof}

\subsection{Proof of Corollary~\ref{coro-winner-loser}}

\begin{proof}
The proof of the claim is by induction on the process that buyers
make the best response bidding. For every such best response
bidding, we use the same proof in the above
Proposition~\ref{prop-winner-winner} and~\ref{prop-loser-loser} to
get the desired result.
\end{proof}


\section{Proofs in Section~4}\label{appendix-proof-max-min}

To prove our results, we need the following definition, which is
similar to $\max$-alternating path.

\begin{definition}
[$\min$-alternating path] Given any equilibrium
$(\mathbf{p},\mathbf{x})$ of a given bid $\mathbf{b}$, let
$G=G(\mathbf{b},\mathbf{p})$ be its demand graph. For any item $j$,
a path $(j=j_1,i_1,j_2,i_2,\ldots,j_\ell,(i_\ell))$ in $G$ is called
a {\em $\min$-alternating path} if edges are not in and in the
allocation $\mathbf{x}$ alternatively, i.e., $x_{i_k}=j_{k+1}$ for
all possible $k$. Denote by
$G_j^{\min}(\mathbf{b},\mathbf{p},\mathbf{x})$ (or simply
$G_j^{\min}$ when the parameters are clear from the context) the
subgraph of $G(\mathbf{b},\mathbf{p})$ (containing both buyers and
items including $j$ itself) reachable from $j$ through
$\min$-alternating paths with respect to $\mathbf{x}$. A
$\min$-alternating path $(j=j_1,i_1,j_2,i_2,\ldots,j_\ell,i_\ell)$
in $G_j^{\min}(\mathbf{b},\mathbf{p},\mathbf{x})$ is called {\em
critical} if $x_{i_\ell}=\emptyset$ and $b_{i_\ell
j_\ell}=p_{j_\ell}$.
\end{definition}

Note that the major difference between $\max$ and $\min$-alternating
paths is that in the former, edges in the path are {\em in} and {\em
not in} the allocation $\mathbf{x}$ alternatively; whereas in the
latter, edges in the path are {\em not in} and {\em in} $\mathbf{x}$
alternatively. Similar to Corollary~\ref{coro-critical-alter-path},
we have the following claim. (The last pair $j_\ell$ and $(i_\ell)$
in a critical $\min$-alternating path is the exact reason that why
the price $p_j$ cannot be decreased further, since otherwise
$i_\ell$ will have to be a winner and items are over-demanded.)

\begin{corollary}\label{coro-critical-min-alter-path}
Given any bid vector $\mathbf{b}$ and
$(\mathbf{p},\mathbf{x})=\mineq(\mathbf{b})$, for any item $j$,
there is a critical $\min$-alternating path in
$G_{j}^{\min}(\mathbf{b},\mathbf{p},\mathbf{x})$.
\end{corollary}

\subsection{Proof of Theorem~\ref{lemma-stable-max=min}}

\begin{proof}
By the definition of $(\mathbf{p},\mathbf{x})$ and
$(\mathbf{q},\mathbf{y})$, we have $\mathbf{p}\le \mathbf{q}$. Let
$T=\{j~|~p_j+\epsilon<q_j\}$ and $S=\{i~|~y_i\in T\}$, i.e., $S$ is
the subset of buyers that win items in $T$ in the \maxeq\
$(\mathbf{q},\mathbf{y})$. Then for any $j\notin T$, either
$p_j=q_j$ or $p_j+\epsilon=q_j$. Assume that $T\neq \emptyset$;
similar to the proof of Lemma~\ref{lemma-basic}, we know that all
buyers in $S$ still win items in $T$ in the \mineq\
$(\mathbf{p},\mathbf{x})$.

Consider any $i_0\in S$ and let $j_0=y_{i_0}\in T$. Consider a bid
vector $(b'_{i_0j})_j$ where $b'_{i_0j_0}=q_{j_0}-\epsilon$ and
$b'_{i_0j}=0$ for any $j\neq j_0$; denote the resulting bid vector
by $\mathbf{b'}$ (where the bids of all other buyers remain the
same). Consider a tuple $(\mathbf{q}',\mathbf{y}')$, where
$q'_j=q_j-\epsilon$ if $j\in T$ and $q'_j=q_j$ if $j\notin T$, and
$y'_i=y_i$ if $i\in S$ and $y'_i=x_i$ if $i\notin S$. It can be seen
that $(\mathbf{q}',\mathbf{y}')$ is an equilibrium for
$\mathbf{b'}$. Note that $b'_{i_0j_0}=q_{j_0}-\epsilon=q'_{j_0}$;
thus, $i_0$ cannot obtain $j_0$ if its price $q'_{j_0}$ is increased
any further.

Let $(\mathbf{q}^*,\mathbf{y}^*)=\maxeq(\mathbf{b'})$; note that
$\mathbf{q}'\le \mathbf{q}^*$. We claim that $i_0$ still wins $j_0$
at price $q^*_{j_0}=q'_{j_0}$ in $(\mathbf{q}^*,\mathbf{y}^*)$.
(This fact implies that $i_0$ obtains more utility from $j_0$ by
paying $\epsilon$ less when bidding $(b'_{i_0j})$; thus \maxeq\
$(\mathbf{q},\mathbf{y})$ is not a Nash equilibrium.) Assume
otherwise, then $i_0$ does not win any item in
$(\mathbf{q}^*,\mathbf{y}^*)$ since $b'_{i_0j}=0$ for any $j\neq
j_0$. Because $q^*_{j_0}\ge q'_{j_0}>p_{j_0}\ge 0$, there must be a
(non-dummy) buyer winning $j_0$: assume that $i_1$ wins
$j_0=y'_{i_0}$, $i_2$ wins $j_1\triangleq y'_{i_1}$, $i_3$ wins
$j_2\triangleq y'_{i_2}$, $\ldots$, $i_k$ wins $j_{k-1}\triangleq
y'_{i_{k-1}}$ in $(\mathbf{q}^*,\mathbf{y}^*)$, and $i_k$ is the
first buyer in the chain that is not in $S$ (such buyer must exist
since $i_0$ is not a winner). Let $j_k\triangleq y'_{i_k}=x_{i_k}$
(note that $j_k\notin T$); then we have $q^*_{j_k}\ge
q'_{j_k}=q_{j_k}\ge p_{j_k}$ and at least one of the two
inequalities is strict (since $i_k$ wins another item $j_{k-1}$ at a
higher price compared to $(\mathbf{p},\mathbf{x})$). This implies
that there must be another buyer $i_{k+1}$ winning $j_k$ in
$(\mathbf{q}^*,\mathbf{y}^*)$. In the process all dummy buyers will
not be introduced to win any item; hence, the same argument
continues and will not stop, which contradicts to the fact that
buyers and items are finite.
\end{proof}

\subsection{Proof of Theorem~\ref{lemma-aligned-price=p*}}

The claim follows from the following two claims.

\begin{proposition}
For any item $j$, $p_j\ge p^*_j$.
\end{proposition}
\begin{proof}
Assume otherwise that there is an item $j_1$ such that
$p_{j_1}<p^*_{j_1}$. Assume that $i_1$ wins $j_1$ in $\mathbf{x}^*$
and $j_2$ in $\mathbf{x}$, i.e., $x^*_{i_1}=j_1$ and
$x_{i_1}=j_{2}$. Since $(\mathbf{p}^*,\mathbf{x}^*)$ is an
equilibrium for the true valuation vector $\mathbf{v}$, we have
$v_{i_1j_1}-p^*_{j_1}\ge v_{i_1j_2}-p^*_{j_2}$. Hence, if
$p_{j_2}\ge p^*_{j_2}$, then
\[v_{i_1j_1}-p_{j_1} > v_{i_1j_1}-p^*_{j_1}\ge v_{i_1j_2}-p^*_{j_2} \ge v_{i_1j_2}-p_{j_2}\]
That is,
$u_{i_1}(\mathbf{p},\mathbf{x})=v_{i_1j_2}-p_{j_2}<v_{i_1j_1}-p_{j_1}
\le d_{i_1}$ (defined in Lemma~\ref{lemma-winner-best-response}). If
$i_1$ has yet made any best response bidding, then he should bid his
best response according to Lemma~\ref{lemma-winner-best-response}.
If he has already made best response bidding, then his bid vector
has already been aligned and the above inequality implies that
$b_{i_1j_1}-p_{j_1}> b_{i_1j_2}-p_{j_2}$, which contradicts to the
fact that $(\mathbf{p},\mathbf{x})$ is an equilibrium for the bid
vector $\mathbf{b}$. Hence, we must have $p_{j_2}< p^*_{j_2}$.

Consider the subgraph $G$ given by the exclusive-OR operation of the
two allocations $\mathbf{x}$ and $\mathbf{x}^*$. For any alternating
path $(i'_1,j'_1,\ldots,i'_\ell,j'_\ell,i'_{\ell+1})$ where $i'_k$
wins $j'_k$ in $(\mathbf{p}^*,\mathbf{x}^*)$ and $i'_{k+1}$ wins
$j'_k$ in $(\mathbf{p},\mathbf{x})$, if $p_{j'_k}<p^*_{j'_k}$ for
any $k$, then by the above argument and considering buyer $i'_k$, we
have $p_{j'_{k-1}}<p^*_{j'_{k-1}}$. Applying the same argument
recursively yields $p_{j'_1}<p^*_{j'_1}$. Since $0\le
v_{i'_1j'_1}-p^*_{j'_1} < v_{i'_1j'_1}-p_{j'_1}$ and $i'_1$ does not
win any item in $(\mathbf{p},\mathbf{x})$, by
Corollary~\ref{coro-best-response} buyer $i'_1$ should continue to
make a best response bidding, which contradicts to the fact that
$(\mathbf{p},\mathbf{x})$ is a Nash equilibrium. Hence, for any
alternating path in $G$, all items have $p_j\ge p^*_j$. For any
alternating cycle in $G$, if there is an item with $p_j<p^*_j$, by
applying the above argument for $j_1$ recursively, all items in the
cycle have $p_j<p^*_j$.

Continue to consider the above item $j_1$ with $p_{j_1}<p^*_{j_1}$.
Since $(\mathbf{p}^*,\mathbf{x}^*)=\mineq(\mathbf{v})$, by
Corollary~\ref{coro-critical-min-alter-path} and abusing notations,
there is a critical $\min$-alternating path
$P=(j_1,i'_2,j'_2,\ldots,i'_{r-1},j'_{r-1},i'_{r})$ in
$G^{\min}_{j_1}(\mathbf{v},\mathbf{p}^*,\mathbf{x}^*)$, where
$x^*_{i'_k}=j'_k$ for $k=2,\ldots,r-1$, $i'_{r}$ does not win any
item and $v_{i'_r j'_{r-1}}=p^*_{j'_{r-1}}$. Consider buyer $i'_2$;
by the definition of $G^{\min}_{j_1}$, we have
\[v_{i'_2j_1}-p_{j_1} > v_{i'_2j_1}-p^*_{j_1} = v_{i'_2j'_2}-p^*_{j'_2}\ge v_{i'_2x_{i'_2}}-p^*_{x_{i'_2}}\]
By the same above argument, we have $p_{x_{i'_2}} < p^*_{x_{i'_2}}$.
Then considering the alternating cycle containing
$j'_2=x^*_{i'_2},i'_2,x_{i'_2}$, we have $p_{j'_2}<p^*_{j'_2}$. The
same argument applies to all items in $P$ recursively; hence, we
have $p_j<p^*_j$ for all items in the path $P$. This implies, in
particular, that $i'_r$ is a winner in $(\mathbf{p},\mathbf{x})$
since $v_{i'_r j'_{r-1}}=p^*_{j'_{r-1}}>p_{j'_{r-1}}$ and by the
best response characterized in Corollary~\ref{coro-best-response}.

Consider the item that $i'_r$ wins in $(\mathbf{p},\mathbf{x})$; let
$j'_r=x_{i'_r}$. Since any buyer in an alternating cycle of $G$ win
in both $(\mathbf{p},\mathbf{x})$ and $(\mathbf{p}^*,\mathbf{x})^*$,
$i'_r$ and $j'_r$ must be at an endpoint of an alternating path of
$G$. Further, by the above argument regarding the prices of items in
alternating paths, we have $p_{j'_r}\ge p^*_{j'_r}$. Therefore,
\[v_{i'_r j'_{r-1}} - p_{j'_{r-1}} > 0 \ge v_{i'_r j'_{r}} - p^*_{j'_{r}} \ge v_{i'_r j'_{r}} - p_{j'_{r}}\]
That is,
$u_{i'_r}(\mathbf{p},\mathbf{x})=v_{i'_rj'_r}-p_{j'_r}<v_{i'_rj'_{r-1}}-p_{j'_{r-1}}
\le d_{i'_r}$ (defined in Lemma~\ref{lemma-winner-best-response}).
We can apply the same argument above for $i_1$ to get a similar
contradiction.

Hence, the claim follows.
\end{proof}

\begin{proposition}
For every item $j$, if $p_j > p^*_j+\epsilon$ then the corresponding
winner can obtain more utility by bidding his best response.
\end{proposition}

\begin{proof}
From the above claim, we have $\mathbf{p}^*\le \mathbf{p}$. Let
$T_1=\{j~|~p^*_j+\epsilon<p_j\}$ and
$T_2=\{j~|~p^*_j+\epsilon=p_j\}$. and $S_1=\{i~|~x_i\in T_1\}$ and
$S_2=\{i~|~x_i\in T_2\}$, i.e., $S_1$ and $S_2$ are the subset of
buyers that win items in $T_1$ and $T_2$ in the \maxeq\
$(\mathbf{p},\mathbf{x})$, respectively. Then for any $j\notin
T_1\cup T_2$, we have $p^*_j= p_j$.

Consider any buyer $i$ and item $j\in T_1\cup T_2$. If $i$ does not
win any item in the \mineq\ $(\mathbf{p}^*,\mathbf{x}^*)$, i.e.,
$x^*_{i}=\emptyset$, then $v_{ij}\le p^*_j < p_j$, which implies
that $i$ cannot win $j$ in $(\mathbf{p},\mathbf{x})$. If $i$ wins an
item not in $T_1\cup T_2$ in $(\mathbf{p}^*,\mathbf{x}^*)$, then
\[v_{ix^*_i}-p_{x^*_i} = v_{ix^*_i}-p^*_{x^*_i} \ge v_{ij}-p^*_{j} > v_{ij}-p_{j}\]
This implies that if $i$ wins $j$ in $(\mathbf{p},\mathbf{x})$, then
he should continue to make a best response bidding by
Lemma~\ref{lemma-winner-best-response}, which contradicts the fact
that $\mathbf{b}$ is a Nash equilibrium. Therefore, all buyers in
$S_1\cup S_2$ win all items in $T_1\cup T_2$ in both
$(\mathbf{p}^*,\mathbf{x}^*)$ and $(\mathbf{p},\mathbf{x})$.

Assume that $T_1\neq \emptyset$; consider any $i_0\in S_1$ and let
$j_0=x^*_{i_0}$. Consider a bid vector $(b'_{i_0j})_j$ where
$b'_{i_0j_0}=p^*_{j_0}+\epsilon$ and $b'_{i_0j}=0$ for any $j\neq
j_0$; denote the resulting bid vector by $\mathbf{b'}$ (where the
bids of all other buyers remain the same). Consider a tuple
$(\mathbf{p}',\mathbf{x})$, where $p'_j=p^*_j+\epsilon$ if $j\in
T_1\cup T_2$ and $p'_j=p^*_j$ if $j\notin T_1\cup T_2$. It can be
seen that $(\mathbf{p}',\mathbf{x})$ is an equilibrium for
$\mathbf{b'}$. Note that $b'_{i_0j_0}=p^*_{j_0}+\epsilon=p'_{j_0}$;
thus, $i_0$ cannot obtain $j_0$ if its price $p'_{j_0}$ is increased
any further.

Similar to the proof of Theorem~\ref{lemma-stable-max=min}, we can
show that in $i_0$ still wins $j_0$ at price $p'_{j_0}$ in
$\maxeq(\mathbf{b'})$. Since
\[v_{i_0j_0}-p'_{j_0}=v_{i_0j_0}-p^*_{j_0}-\epsilon \ge v_{i_0x_{i_0}}-p^*_{x_{i_0}}-\epsilon >  v_{i_0x_{i_0}}-p_{x_{i_0}}\]
we know that $i_0$ obtains more utility when changing bids from
$\mathbf{b}$ to $\mathbf{b}'$, which is again a contradiction.
\end{proof}

\subsection{Proof of Theorem~\ref{theorem-social-welfare}}

\begin{proof}
Consider the equilibrium
$(\mathbf{p},\mathbf{x})=\maxeq(\mathbf{b})$. For any loser $i$, by
the best response rule of Lemma~\ref{lemma-loser-best-response}, we
know that $v_{ij}\le p_j$ for all items. For any winner $i$, we have
$v_{ix_i}-p_{x_i}\ge 0$; further, for any item $j\neq x_i$, by the
best response rule of Lemma~\ref{lemma-winner-best-response}, we
have $u_i(\mathbf{p},\mathbf{x})=v_{ix_i}-p_{x_i}\ge d_i\ge
v_{ij}-p_{j}$, where the first inequality follows from the fact that
the dynamics converges, and the second inequality follows from the
observation that when $i$ bids zero, all prices will not increase.

Let $\mathbf{x}^*$ be an efficient allocation and maximizes social
welfare. Given the existence of dummy buyers, we can assume without
loss of generality that all items all sold out in both $\mathbf{x}$
and $\mathbf{x}^*$. Consider the exclusive-OR relation given by the
two allocations; it can be seen that it contains either alternating
paths of even length where both endpoints are buyers or alternating
cycles.

If there is an alternating path of even length, say,
\[i_1 \stackrel{\mathbf{x}}{\textup{-----}} j_1 \stackrel{\mathbf{x}^*}{\textup{-----}} i_2 \stackrel{\mathbf{x}}{\textup{-----}} j_2 \stackrel{\mathbf{x}^*}{\textup{-----}} \cdots \cdots \stackrel{\mathbf{x}^*}{\textup{-----}} i_{r-1} \stackrel{\mathbf{x}}{\textup{-----}} j_{r-1} \stackrel{\mathbf{x}^*}{\textup{-----}} i_r\]
By the above discussions, we have
\begin{eqnarray*}
v_{i_1j_1}-p_{j_1} &\ge& 0 \\
v_{i_{k}j_{k}}-p_{j_k} &\geq& v_{i_k j_{k-1}}-p_{j_{k-1}}, \ 2 \leq k \leq r-1 \\
0 &\ge& v_{i_{r}j_{r-1}} - p_{j_{r-1}}
\end{eqnarray*}
Adding these inequalities together yields
\[\sum_{(i,j)\in \mathbf{x}}v_{ij} = \sum_{1 \leq k \leq r-1}{v_{i_{k}j_{k}}} \geq \sum_{2 \leq k \leq r}{v_{i_kj_{k-1}}} = \sum_{(i,j)\in \mathbf{x}^*}v_{ij}\]
which implies that $\mathbf{x}$ has the same maximum total valuation
on such a path as $\mathbf{x}^*$.

The analysis is similar on alternating cycles. Hence, the allocation
$\mathbf{x}$ is efficient as well, which completes the proof.

\end{proof}


\section{Nash Equilibria with Various Equilibrium Prices}\label{appendix-example-nash}

The following examples show that in general the \maxeq\ prices in a
Nash equilibrium can be either (much) smaller or higher than the
\mineq\ price vector at truthful bidding. These examples are in
contrast with the statement of Theorem~\ref{lemma-aligned-price=p*}
which says that when buyers bid aligned vectors initially, we always
have $\maxeq(\mathbf{b})\approx \mineq(\mathbf{v})$.

\begin{example}\label{example-nash-less-min}
{\sc (Nash equilibrium with small \maxeq\ prices)} There are $n+1$
buyers $i_0,i_1,\ldots,i_n$ and $n$ items $j_1,\ldots,j_n$. Buyer
$i_0$ only desires $j_1$ with value $v_{i_0j_1}=n\epsilon$; buyer
$i_k$ only desires $j_k$ and $j_{k+1}$ with value
$v_{i_kj_k}=v_{i_kj_{k+1}}=n\epsilon$ for $k=1,\ldots,n-1$; and
buyer $i_n$ only desires $j_n$ with value $v_{i_nj_n}=n\epsilon$.
That is, all buyers have the same value $n\epsilon$ for the items
that they desire. It can be seen that in the minimum equilibrium
$(\mathbf{p}^*,\mathbf{x}^*)=\mineq(\mathbf{v})$, $p^*_j=n\epsilon$
for all items. On the other hand, consider the bid vector
$\mathbf{b}$, allocation vector $\mathbf{x}$ (where
$x_{i_k}=j_{k+1}$ and $i_n$ does not win any item), and price vector
$\mathbf{p}$ given by the following figure:
\begin{figure}[ht]
\begin{center}
\includegraphics[scale = 0.9]{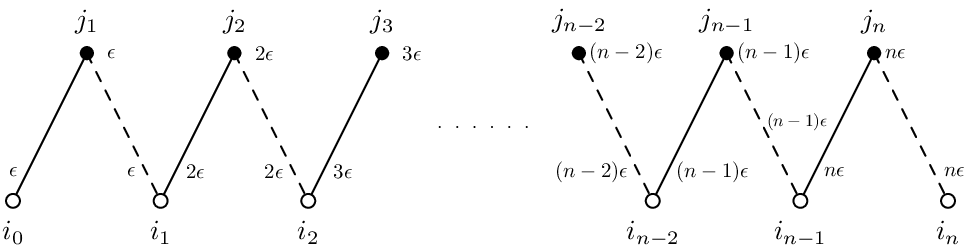}
\end{center}
\end{figure}

\noindent It can be seen that $(\mathbf{p},\mathbf{x})$ is a \maxeq\
of $\mathbf{b}$. Further, $\mathbf{b}$ is a Nash equilibrium for the
\maxeq\ mechanism. (Indeed, for any buyer $i_k$, the maximum
equilibrium price vector for bid vector $\mathbf{b}_{\not\ni i_k}$
where $i_k$ bids zero for all items is the same as $\mathbf{p}$. For
any buyer $i_k$, $1\le k \le n-1$, by
Lemma~\ref{lemma-winner-best-response}, the utility of $i_k$ is
maximized when bidding $b'_{i_kj_k}=b'_{i_kj_{k+1}}=(k+1)\epsilon$;
in this case, however, the price of $j_k$ will be increased to
$(k+1)\epsilon$; hence, $i_k$ will obtain the same utility as
$u_{i_k}(\mathbf{p},\mathbf{x})$. We can verify that buyers $i_0$
and $i_n$ cannot obtain more utilities similarly.) Hence, the
equilibrium price vector in a Nash equilibrium of the \maxeq\
mechanism can be (much) smaller than the minimum equilibrium price
vector $\mathbf{p}^*$.
\end{example}

\begin{example}\label{example-nash-less-min}
{\sc (Nash equilibrium with large \maxeq\ prices)} There are $n+1$
buyers $i_0,i_1,\ldots,i_n$ and $n$ items $j_1,\ldots,j_n$. Buyer
$i_0$ only desires $j_1$ with value $v_{i_0j_1}=\epsilon$; buyer
$i_k$ only desires $j_k$ and $j_{k+1}$ with value
$v_{i_kj_k}=v_{i_kj_{k+1}}=n\epsilon$ for $k=1,\ldots,n-1$; and
buyer $i_n$ only desires $j_n$ with value $v_{i_nj_n}=n\epsilon$. It
can be seen that in the minimum equilibrium
$(\mathbf{p}^*,\mathbf{x}^*)=\mineq(\mathbf{v})$, $p^*_j=\epsilon$
for all items. On the other hand, consider the bid vector
$\mathbf{b}$, allocation vector $\mathbf{x}$ (where $x_{i_k}=j_{k}$
and $i_0$ does not win any item), and price vector $\mathbf{p}$
given by the following figure:
\begin{figure}[ht]
\begin{center}
\includegraphics[scale = 0.9]{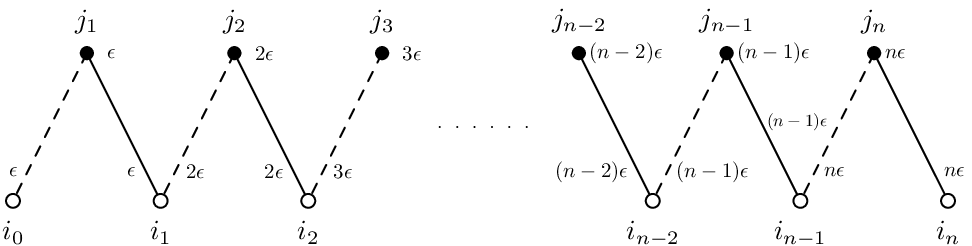}
\end{center}
\end{figure}

\noindent It can be seen that $(\mathbf{p},\mathbf{x})$ is a \maxeq\
of $\mathbf{b}$. Further, $\mathbf{b}$ is a Nash equilibrium for the
\maxeq\ mechanism. (Indeed, for any buyer $i_k$, the maximum
equilibrium price vector for bid vector $\mathbf{b}_{\not\ni i_k}$
where $i_k$ bids zero for all items is the same as $\mathbf{p}$.
Similar to the above example, we can verify that no buyer cannot
obtain more utility by unilaterally changing his bid.)
Hence, the equilibrium price vector in a Nash equilibrium of the
\maxeq\ mechanism can be (much) larger than the minimum equilibrium
price vector $\mathbf{p}^*$.
\end{example}

The above examples are not contradictory to
Theorem~\ref{lemma-aligned-price=p*}, which says that starting from
any aligned bid vector, the aligned best response dynamics converges
to the \mineq\ prices. In particular, the initials bid vectors
$\mathbf{b}$ in the above examples are not aligned; hence, the claim
in Theorem~\ref{lemma-aligned-price=p*} does not apply here.